\documentclass[conference,compsoc]{IEEEtran}
\usepackage[cmex10]{amsmath}
\usepackage{array}
\usepackage{url}
\usepackage{times}
\usepackage{listings}
\usepackage[dvipsnames,table]{xcolor}
\usepackage{dsfont}

\newcommand{\acl}[0]{\textsc{ACL2}}

\newcommand{\ie}{\emph{i.e.}}
\newcommand{\eg}{\emph{e.g.}}

\newcommand{\R}[0]{\ensuremath{\mathcal{R}}}

\usepackage{makecell}

\newcommand{\bc}{BitCoin} 
\newcommand{\gs}{GossipSub}

\newcommand{\fc}{FileCoin}
\newcommand{\Eth}{Eth2.0}
\newcommand{\OldEth}{Eth1.0}
\newcommand{\score}{score function}
\newcommand{\Score}{Score Function}
\newcommand{\acls}{ACL2s}
\newcommand{\golang}{Golang}

\newcommand{\BLOCKS}{\texttt{BLOCKS}}

\newcommand{\AGG}{\texttt{AGG}}
\newcommand{\SUBONE}{\texttt{SUB1}}
\newcommand{\SUBTWO}{\texttt{SUB2}}
\newcommand{\SUBTHREE}{\texttt{SUB3}}

\newcommand{\group}{\texttt{Group}}
\newcommand{\p}{\texttt{peer-id}}

\newcommand{\ps}{\texttt{peer-state}}
\newcommand{\nts}{\texttt{nbr-topic-state}}
\newcommand{\mst}{\texttt{msgs-state}}
\newcommand{\tctrs}{\texttt{topic-counters}}
\newcommand{\gctrs}{\texttt{global-counters}}
\newcommand{\nbrscores}{\texttt{nbr-scores}}

\newcommand{\evnt}{\texttt{evnt}}
\newcommand{\hbm}{heartbeat maintenance}
\newcommand{\twp}{\texttt{twp}}

\newcommand{\mmd}{\texttt{meshMessageDeliveries}}

\newcommand{\trc}{\texttt{trace}}
\newcommand{\ptt}{\texttt{pt-tctrs-map}}
\newcommand{\fin}{\textbf{F}}
\newcommand{\glb}{\textbf{G}}

\newcommand{\blue}{\mbox{blue}}

\newcommand{\ind}{indicator}
\newcommand{\inds}{indicators}
\newcommand{\Inds}{Indicators}

\newtheorem{property}{Property}
\newcommand{\ag}[1]{$\mathrm{AG}_{#1}$}

\newcommand{\msgs}{\mathit{Msgs}}
\newcommand{\events}{\mathit{Events}}

\newcommand{\hbme}{\mathit{H}}

\makeatletter
\let\sv@thm\@thm
\def\@thm{\let\indent\relax\sv@thm}
\makeatother

\usepackage{tikz}
\usetikzlibrary{positioning, shapes.geometric, backgrounds}
\usepackage{tikzscale}

\usepackage{mathtools}

\usepackage{enumerate}
\usepackage[shortlabels]{enumitem}

\usepackage[hidelinks]{hyperref}
\usepackage{breakurl}

\usepackage{physics}
\usepackage{tikz}
\usepackage{mathdots}
\usepackage{yhmath}
\usepackage{cancel}
\usepackage{color}
\usepackage{siunitx}
\usepackage{array}
\usepackage{multirow}
\usepackage{amssymb}
\usepackage{gensymb}
\usepackage{tabularx}
\usepackage{extarrows}
\usepackage{booktabs}
\usetikzlibrary{fadings}
\usetikzlibrary{patterns}
\usetikzlibrary{shadows.blur}
\usetikzlibrary{shapes}

\usepackage{adjustbox}
\usepackage{float}
\usepackage{pgfplots}
\pgfplotsset{compat=1.8}

\usepackage{derivative}

\pgfplotsset{ 
  MyQuiver2D/.style={
    width=0.6\textwidth, 
    axis equal image, 
    view={0}{90}, 
    xmin=-2.1, xmax=2.1, 
    ymin=-2.1, ymax=2.1,
    domain=-2:2, y domain=-2:2, 
    xtick={-2,-1.5,...,2}, ytick={-2,-1.5,...,2}, 
    samples=21, 
    cycle list={    
      gray,
      quiver={
        u={f(x,y)}, v={g(x,y)}, 
        scale arrows=0.015,
        every arrow/.append style={
          -latex 
        },
      }\\
      red, samples=31, smooth, very thick, no markers, domain=-2:2\\ 
    }
  }
}

\usepackage[linesnumbered,ruled,vlined]{algorithm2e}
\begin{document}

\title{Formal Model-Driven Analysis of Resilience of GossipSub to Attacks from Misbehaving Peers}

\author{\IEEEauthorblockN{Ankit Kumar$^\diamond$,
 Max von Hippel$^\diamond$,
 Panagiotis Manolios$^\dagger$,
 Cristina Nita-Rotaru$^\dagger$}

\IEEEauthorblockA{Northeastern University, Boston, MA, USA\\
 \{kumar.anki,vonhippel.m,p.manolios,c.nitarotaru\}@northeastern.edu}}

\date{}

\maketitle
\def\thefootnote{$\diamond$}\footnotetext{Contributed equally.}
\def\thefootnote{$\dagger$}\footnotetext{Listed alphabetically.}
\renewcommand{\thefootnote}{\arabic{footnote}}


\begin{abstract}
\gs~is a new peer-to-peer communication protocol
designed to counter attacks
from misbehaving peers by controlling what information
is sent and to whom,
via a
\emph{\score} computed by each peer that captures positive and negative
behaviors of its neighbors.
The score function depends on several parameters (weights,
caps, thresholds) that can be configured by applications using \gs.
The
specification for \gs~is written in English and its resilience to
attacks from misbehaving peers is supported
empirically by emulation testing using an
implementation in \golang.
  
In this work we take a foundational approach to understanding the
resilience of \gs~to attacks from misbehaving peers.
We build the first formal model of
  \gs, using the \acls\ theorem prover.  Our model is officially
  endorsed by the \gs\ developers. It can simulate \gs~networks of
arbitrary size and topology, with arbitrarily configured peers, and
can be used to prove and disprove theorems about the protocol.  We
formalize fundamental security properties stating that the \score\ is
fair, penalizes bad behavior, and rewards good behavior.  We prove that
the \score\ is always fair, but can be configured in ways that either
penalize good behavior or ignore bad behavior.  Using our model, we
run \gs~with the specific configurations for two popular real-world
applications: the \fc~and \Eth~blockchains. We show that all
properties hold for \fc. However, given any \Eth\ network (of any
topology and size) with any number of potentially misbehaving peers,
we can synthesize attacks where these peers are able to continuously
misbehave by never forwarding topic messages, while maintaining
positive scores so that they are never pruned from the network by \gs.
\end{abstract}


\section{Introduction}
\label{sec:intro}

Gossip protocols are a fundamental building block for communication in
peer-to-peer (P2P) systems.  One such protocol is \gs, and is used by popular applications 
such as \fc~\cite{FCwhitepaper} and \Eth~\cite{EthWhitepaper}. 
As of November 2022, the market cap of \fc\ is \$1.4B USD~\cite{fcCap}.
\Eth\ is the second most valuable cryptocurrency, after 
\bc, with a November 2022 market cap of \$143B USD~\cite{ethCap}.
A major concern in P2P systems including \gs\ 
is the risk of 
attacks launched by misbehaving peers.
As such attacks can have significant financial implications, 
it is critical that \gs\ is resilient to
attacks from misbehaving peers.

\gs\ addresses attacks by misbehaving peers via fine-granularity dissemination
techniques and heuristic defense mechanisms
that are carefully controlled by a \score\ capturing positive and negative peer behaviors,
globally and within topics. (\gs\ partitions communication by
  \emph{topics} to which peers can subscribe or unsubscribe
  -- as in pub/sub systems.) 
Scores calculated using the
  \score\ are local to a peer and are not shared with other peers.
Several aspects of the \score\ are crucial to its
correctness, e.g.: defining good and bad behavior, specifying
weights that are applied to good and bad behavior in the \score, and
choosing thresholds for making decisions about peer behaviors, to
name just a few. Applications using \gs\ can configure
such parameters.

{\em Where does the assurance that \gs\ is indeed resilient to
  attacks from misbehaving peers come from?} \gs\ is defined by a prose
specification~\cite{gs1.0,gs1.1} and an implementation in
\golang~\cite{golanggs}.  The \golang\ implementation was
subjected to unit tests and manual code review by expert
programmers, and tested under various threat models at scale
using the \textsc{Testground} network
emulator~\cite{vyzovitis2020gossipsub,leastAuth20}.

While the manual reviews and emulations empirically suggest that the
protocol is resilient to attacks, there is no rigorous
specification of what this resilience means or of what assumptions are
necessary for the correctness of the system. Formal methods can help
disambiguate system specifications and formulate implicit assumptions
made by the system designers. They also can expose flaws in system
requirements, often not captured through testing.  In contrast to
testing, formal methods provide mathematical proofs that show that a
system does or does not behave correctly.

Formal methods have been previously applied to gossip
and pub/sub protocols~\cite{garlan2003model,baresi2007accurate,kwiatkowska2008analysis,diaz2015model,oxford2020quantitative}, however with
important limitations.  First, they studied much simpler protocols
than \gs, based on \emph{flooding}, where data is disseminated naively
across the network without regard for the bandwidth.
Second, they used model checking and made simplifying assumptions to
avoid state-space explosion.

\textbf{Our contributions.}
We focus on rigorously studying \gs\ and its resilience
to attacks from misbehaving peers using \acls, a
theorem prover based on a purely functional \textsc{Lisp}-based
language~\cite{DillingerMVM07,acl2s11}. \acls\ is highly expressive,
allowing us to express arbitrary computation and properties over
infinite-state systems. Our contributions
are:

\noindent {\bf $\bullet$~A formal model of \gs:} In contrast to prior works that
  studied gossip protocols using heavily simplified or restricted
  models, we study \gs\ by modeling every aspect of its prose
  specification. Our model is not just an abstraction, it is an {\em
    actual executable program} that can be formally reasoned about.
  When we find the specification ambiguous, we compare it to
  the implementation and also consult the specification authors. Our
  model can simulate \gs\ networks of arbitrary size and topology,
  with arbitrarily configured peers, and can be used to prove or
  disprove protocol properties.  It is publicly available at \url{github.com/gossipsubfm},
  allowing developers of applications using \gs\ to
  verify security properties for the configuration corresponding to
  their application, and was officially endorsed
  by the \gs\ developers as a formal specification for the protocol in
  their documentation~\cite{gsreadme}.

\noindent {\bf $\bullet$~Security properties and analyses:} Since the prose
  specification~\cite{gs1.0,gs1.1} and emulation
  analysis~\cite{vyzovitis2020gossipsub} do not list
  properties, we formalize four security properties about the \score\
  that can be inferred based on a close reading of these
  documents. These are necessary for the score-based
  defense mechanisms to defend against attacks from
    misbehaving peers.
\begin{enumerate}[leftmargin=*,label=(\arabic*)]
\item If a peer's performance for some topic is continuously
  non-positive, then, eventually, the peer's score will be
  non-positive.
    \item When a peer misbehaves, its score decreases.
    \item When a peer behaves, its score does not decrease.
    \item Peers are scored fairly: if they appear to behave
      identically, they are given identical scores.
\end{enumerate}
We prove that (\ref{p3}) and (\ref{p4}) hold for all \gs\ configurations.
In contrast, using \acls, we automatically find configurations for
which (\ref{p1}) and (\ref{p2}) fail. We prove that the configuration
used by \Eth\ is one such configuration where both these properties
fail, and we prove that the \fc\ configuration satisfies all four
although it achieves this by compromising important
protocol functionality.

\noindent {\bf $\bullet$~Attack generation:} We show
how violations of these properties for \Eth\ can be used to create
attacks against the entire network.  Our attacks
  exploit the fact that the \score\ can be configured in ways where
  peers can misbehave without penalty.  \Eth\ uses one such
  configuration.  To find these attacks we formalize what it
  means for the protocol to behave correctly, and then ask the
  \acls\ theorem prover if it was possible for the protocol to behave
  incorrectly. 
In contrast, prior emulation and expert code review of \gs\ only looked at specific pre-programmed attack scenarios, \eg, where an honest peer is surrounded by malicious peers who delay or drop messages forwarded from the honest peer, or where the network is saturated with malicious peers who instantaneously stop forwarding data.  We take a more general approach, formalizing what it means for \gs\ to behave correctly, and then asking whether any attack scenarios exist -- including unknown ones -- in which the protocol might behave incorrectly.
   We synthesize and verify attacks violating the first
property for \Eth. These attacks can be carried out on any
\Eth\ network, regardless of the topology or size, and allow
peers to continuously misbehave, by never forwarding messages in target topics, 
while maintaining
positive scores so that they are never pruned from the network by the \gs\
layer of \Eth.
Finally, we also show that \fc\ uses \gs\ configurations that
violate the \gs\ specification.

{\bf Ethics.} We submitted responsible disclosures to the \gs\ developers at
Protocol Labs, as well as the Ethereum Foundation.  Both groups
provided feedback, agreeing with our results.  The Ethereum Foundation
is working on a patch, and notified maintainers of popular Ethereum
implementations about the issue.  An alternative to waiting for a
patch is to use flooding at the cost of increased network consumption,
which \gs\ was designed to avoid.


\section{Background}
\label{sec:bk}
We provide background on gossip protocols 
and attacks against them. We then overview previous 
work applying formal methods to gossip protocols
and describe our approach. 

\subsection{Gossip Protocols and Misbehaving Nodes}
\label{sec:bk_gp}
P2P systems construct logical networks without requiring peers to maintain information 
about the global topology of the P2P network. Peers maintain information about their neighbors, peers
they can communicate with directly. Communication between peers that
are not neighbors is achieved through gossip protocols that propagate
information throughout the network by having each peer disseminate
information using its local information about other peers. 

P2P systems are engineered to deal with not only system dynamics, 
such as {\em churn}  where peers join and leave the system as desired, 
peer failures, and network partitions,  
but also with attacks from misbehaving peers. 
Such misbehaving peers can be Sybils, or peers that have
been compromised by an attacker. In Sybil attacks, a
single attacker orchestrates a multitude of identities (called Sybils)
to gain unfair influence over the network~\cite{douceur2002sybil}.  In
the absence of a central entity for authentication, defenses against
Sybils have focused on examining the network topology and
looking for anomalies in this graph.  In many real systems such solutions
requiring global information are impossible, so more local
approaches were proposed, \eg, examining the geo-location of IP
addresses (this is not a robust defense, considering how easy is to
fake IP addresses), or imposing a network topology that constrains an
attacker in what identity they can assume in the system. More recently,
some systems focused on the functionality of the application itself
and made acting as part of the system incur a computation cost
(proof-of-work) -- as in \bc, where
the constraint is computational, and the correctness of the system
relies on assumptions about the computational power available to the
attacker and the theoretical complexity of the proof-of-work problem.
Recent solutions against Sybil attacks were also proposed for 
social \cite{sybilguard_sigcomm2006,sybillimit_sp2008,sybilinfer_nsdi2009,sybil_ccs2019} 
and vehicular \cite{sybil_vn2013,sybil_vn2014} networks, where such attacks are
also prevalent.

In gossip protocols, the main impact that misbehaving
nodes can have is to disrupt communication by 
dropping or delaying application data or P2P control messages.
Gossip protocols that do
not use flooding are more vulnerable to attacks from misbehaving nodes
as a small number of nodes can disrupt communication, potentially across
the entire system. Since peers are expected to deliver application and
control messages, maintain the logical network, and signal operational
status, a potential defense against misbehaving nodes is to observe
peer behavior and use this information to decide to whom new messages
should be forwarded.

\subsection{Formal Methods for Gossip Protocols}
\label{sec:bk_fm}
  
Formal methods (FM) refer to tools and techniques used to specify and
reason about systems with mathematical rigor, using logic.
Mathematical specifications of systems can be used to formalize all
possible system behaviors as well as properties that systems are
expected to satisfy.  There are many formal techniques for either
proving or disproving that systems,  
or abstractions thereof, satisfy properties.  One class of tools,
which includes interactive theorem proving, has high expressiveness,
allowing one to specify arbitrary, Turing-complete computational
systems and properties.  Such tools require well-trained human
proof engineers with the ability to interact with the tools in order
to obtain formal, mechanically-checked proofs.  Another class of tools
includes decision procedures for restricted fragments of logic, such
as temporal logic.  Such tools can be used in a more automated way,
although they impose severe limitations on what can be expressed, \eg:
properties over integers with only addition, multiplication, and
equality, due to their undecidability, are already too expressive to
be handled by such tools. Examples of such decision procedures include
automated theorem provers (such as SMT solvers), model checkers, type
checkers, and static analyses based on abstract interpretation. Using
decision procedures effectively often requires reasoning about
expressible abstractions of systems, \eg, abstractions based on types
or abstract domains or finite-state
abstractions~\cite{blanchet2016modeling,meier2013tamarin}.

{\bf Prior works applying FM to gossip protocols.}
Multiple prior works proposed general frameworks for verifying pub/sub protocols, but did not consider security or attacks~\cite{garlan2003model,baresi2007accurate,he2007formal}.
In a similar vein, D{\'\i}az~et.~al. model-checked a pub/sub architecture for discoverable web services~\cite{diaz2015model}.
Dagand~et.~al. created \textsc{Opis}, an \textsc{OCaml} framework for building and reasoning about distributed systems, which included a formal framework for defining gossip protocols.  Systems built in their tool can be evaluated using the \textsc{Isabelle} and \textsc{Coq} theorem provers, or using a model checker and simulator of their own design~\cite{dagand2009opis}.
Bakhshi~et.~al. 
surveyed formal methods techniques that could be applied to gossip protocols~\cite{bakhshi2007formal}.
A subsequent work
built a pen and paper framework for modeling dissemination in gossip protocols by abstracting their behaviors to just pair-wise 
interactions~\cite{bakhshi2011modeling}.
Van Ditmarsch~et.~al. built an epistemic model checker as part of their framework to improve dissemination in gossip protocols~\cite{van2019strengthening}.
Two prior works studied gossip protocols using 
probabilistic model checking~\cite{kwiatkowska2008analysis,oxford2020quantitative}.
Because of state-space explosion, all the model checking
papers had to abstract the protocol logic and/or
restrict the properties they studied.

Gossip protocols that were previously studied 
with formal methods used (partial or total) flooding,
so that even if misbehaving nodes decide not to forward data, in a sufficiently
well connected network, every message will eventually reach every
node. Thus, prior works that applied formal methods to gossip
protocols focused on proving that all
messages were eventually fully disseminated.  This approach does not
apply for protocols that balance bandwidth overhead with data delivery
(such as \gs) as they have different specifications and safety
properties.

\subsection{Our Approach}
\label{ss::approach}
We study \gs,
a gossip service 
that addresses attacks from misbehaving nodes by using a score function to capture peer behavior
combined with defense mechanisms that adaptively modify the local
network topology.
We use interactive theorem proving because methods based on decision
procedures, such as model checking, cannot be used to study the
actual infinite-state protocol.  While theorem proving
requires more human effort, it allows us to provide a formal,
executable model of the protocol, to formalize properties that
the protocol should satisfy, and to prove or disprove such properties
for various configurations.

Note that in an interactive theorem-prover, we can articulate any
  Turing machine (including infinite-state systems) and any predicate
  logic property about it, but, the prover might not be able to prove
  or disprove the property without expert human guidance.  In
  contrast, symbolic model-checkers like
  \textsc{Spin}~\cite{holzmann1997model},
  \textsc{Tamarin}~\cite{meier2013tamarin}, and
  \textsc{ProVerif}~\cite{blanchet2016modeling},
  only support restricted models and logics that lend themselves to
  automated analysis.
  For example, \textsc{Spin} supports finite Kripke Structures and
  Linear Temporal Logic properties, while \textsc{ProVerif} supports
  an applied $\pi$-calculus with cryptographic primitives, and
  properties relating to secrecy, authentication, and
  process-equivalence.

We use the \acl\ Sedan
(\acls)~\cite{dillinger-acl2-sedan,chamarthi2011acl2} theorem prover, which extends
\acl~\cite{acl2-car, acl2-web} with an advanced data definition
framework (\emph{Defdata})~\cite{defdata}, the \emph{cgen} framework
for automatic counterexample generation~\cite{chamarthi-integrating-testing,harsh-fmcad,harsh-dissertation},
a powerful termination analysis based on calling-context
graphs~\cite{ccg} and ordinals~\cite{ManoliosVroon03, ManoliosVroon04,
  MV05}, a property-based modeling/analysis framework, and IDE
support.

In contrast to other theorem-provers, \acls\ allows us to build
  an \emph{executable} model using the Defdata framework, and then
  generate attack specifications against that model using
  the cgen framework -- which rivals or out-performs other
  state of the art tools such as \textsc{Alloy} or
  \textsc{LEAN}'s \textit{hammer} tactic~\cite{harsh-fmcad}.  And since \acls\ is
  \textsc{Lisp}-based, the model is more expressive and readable
  to the average software engineer than, \eg, \textsc{Coq} or
  \textsc{LEAN} code.  Reasoning in
  \acls\ is facilitated by a collection of proof methods including rewriting, numerous decision procedures, and
a large collection of libraries.  Thus, we model \gs\ by
implementing it as a fully functional computer program in \acls, and
then we reason about it, all in the same system.

Our model can be used for large-scale simulations, as a formal
specification for \gs, and also as a reference with which to
prove or disprove properties of \gs.
To the best of our knowledge, we are the first to fully formalize an executable model
of a non-trivial gossip protocol not entirely based on flooding, and 
then automatically prove and disprove properties about that protocol.


\section{\gs}\label{sec:gs}
In this section, we overview the design of \gs,
provide more details about the \score,  and describe how \gs\ was validated by its
designers.

\subsection{Overview}\label{subsec:gs-overview}

The basic approach to quickly disseminate information in a P2P system is
where every peer forwards every new message to all of its neighbors,
\emph{flooding} the network. Because data travels on
all possible paths in the P2P network, this approach is the most resilient to
 attacks from misbehaving nodes that do not
 correctly forward messages.
However, all this dissemination incurs a
significant bandwidth cost.

\gs\ was proposed to decrease this bandwidth cost by using a mechanism
called \emph{lazy pull} to balance speed of message dissemination with
bandwidth consumption. Specifically, the metadata of
messages are periodically disseminated in a controlled manner, whereas
full messages are sent upon request.  \gs\ partitions 
data in \emph{topics} to which
peers can subscribe or unsubscribe as in pub/sub systems.  For
each topic, nodes create and maintain a dissemination
topology.  If the node subscribes to the topic, the topology is
called a \emph{peer mesh}, otherwise it is a
\emph{peer fanout}.  A peer's meshes and fanouts are subsets of
  its peer-list, and the mesh and fanout for a given topic are
  disjoint.

Unfortunately, by avoiding flooding, \gs\ becomes less resilient to
attacks against communication from malicious
nodes. In such attacks, malicious nodes either do
not forward data, or do so on a delayed schedule.  To address
this, \gs\ uses a set of defense mechanisms based on a
{\em score} that is locally maintained by each peer for each of its
neighbors, capturing their observable positive and negative behaviors.
A positive/negative score is intended to indicate {\em
  good}\emph{/bad} behavior, respectively.  Peers
re-calculate scores periodically and use them to adjust their meshes
and fanouts, determining to whom they will send
data.

\begin{figure}
\centering
\includegraphics[width=\columnwidth]{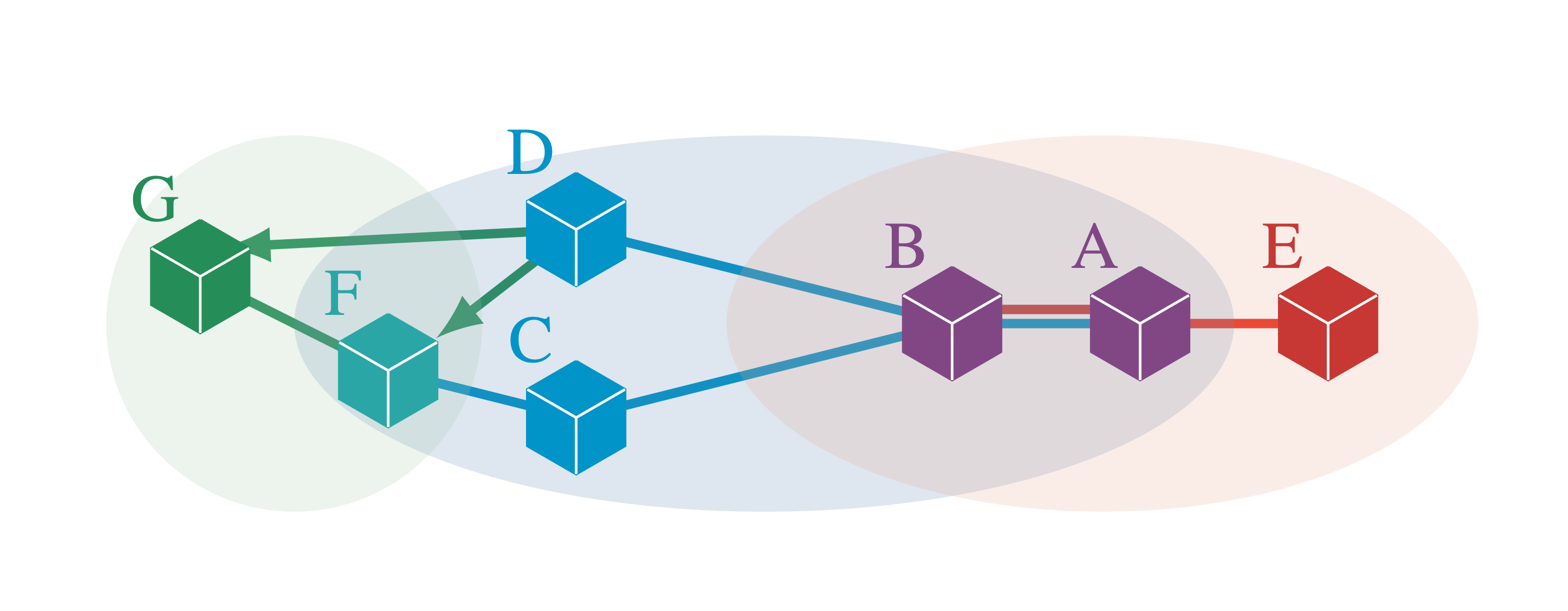}
\caption{Example \gs~network where cubes denote peers, each
  ellipse contains all the peers that subscribe to a particular
  (colored) topic.
  }
\label{fig:mesh-fanout-diagram}
\end{figure}
  
\noindent$\bullet$~{\bf Message dissemination.}  Each peer~$p$ is initialized with a mutable list of other
peers and their subscriptions -- these listed peers are 
the \emph{neighbors} of~$p$.  Over time new neighbors can join and existing
neighbors can leave the network.  
Peers and their neighbors
communicate over topics.  We denote by $p.T$ the set of topics peer~$p$ is
aware of and by $p.S$\ the set of topics that $p$ subscribes
to. Both sets are mutable and we define $p.U = p.T \setminus p.S$ to
be the topics to which~$p$ does not subscribe.  
Each peer~$p$ communicates
full-messages on a subscribed topic~$s$ only to a subset of
its $s$-subscribing neighbors, denoted~$p.M(s)$ and called a \emph{mesh}.
Likewise, a peer~$p$ communicates full-messages on a topic $u$ to which it 
does not subscribe only to a subset of its $u$-subscribing peers~$p.F(u)$, 
called a \emph{fanout}.  Meshes and fanouts are mutable, as are subscriptions, 
meaning a peer might unsubscribe from a topic, 
delete its corresponding mesh, and build a corresponding fanout; or vice versa.

\noindent$\bullet$~{\bf Metadata dissemination.}
Metadata about recently received mesh and fanout messages are periodically broadcast to a
newly randomly selected subset of peers, allowing
metadata to disseminate quickly with low overhead,
so peers can request specific messages from
whoever has that content.

\noindent$\bullet$~{\bf \gs\ threat model.}
The \gs\ developers assume the following (implicitly or explicitly): applications frequently inject new messages for dissemination; every network peer runs the same application with the same configuration; and the goal of honest peers is the rapid, on-demand, total dissemination of uncorrupted data, with low overhead.  
Honest peers follow the \gs\ state machine, responding to requests and forwarding data as quickly as possible, whereas malicious ones can perform any of the following network actions: sending valid or invalid messages, forwarding data with any amount of delay, dropping data to be forwarded, or sending any \gs\ control message at any time.  The goal of the malicious peers is to misbehave by dropping or delaying data forwarding or by sending invalid messages, without their malicious actions being detected.

\noindent$\bullet$~{\bf Defense mechanisms.}  \gs\ restricts the mesh and fanout peers
to only those who appear less likely to be malicious
nodes.  This determination is made based on a \score\ that each peer
computes about each of its neighbors.  The \score\ is used to
remove (prune) and add (graft) peers, \eg,
in Fig.~\ref{fig:mesh-fanout-diagram}, if peer $A$ penalizes $B$ for
sending invalid messages, causing $B$'s score
to become negative, then $B$ will be pruned from $A$'s meshes.

\subsection{The \Score}\label{subsec:score}

\noindent{\bf Peer behavior.}  The goal of the \score\ is to measure good and
bad behaviors of peers.  At a high level, the \score\ takes as input a
list of counters, 
that count specific good and bad behaviors of the peer being scored. 
Some of these counters are indexed by topic (and are called topic-specific) 
while others are not (and are called global).
For example, the 
invalid message deliveries counter 
for a neighbor~$q$
on a topic~$t$ counts the number of invalid message deliveries the scoring peer~$p$
has received from~$q$ on the topic~$t$.
Some of the counters decay
 periodically, so that recent events influence the score more than
 historical ones, and the degree to which each counter decays is specific to the 
 counter (and topic, if the counter is topic-specific)
 and configured by the application.
These counters are used to compute
\inds~that measure good and bad behaviors of the peer being scored, 
such as, the \ind~$P_6(t)$
which equals the invalid message deliveries, squared.
Such \inds\ might be non-continuous, and might take as input more than one
counter.  Like the counters, some of the \inds\ are topic-specific while others
are global.

The \score\ multiplies the good behavior \inds\ with
positive weights, and the bad behaviors \inds\ with
negative weights, and then combines the resulting values using
weighted and capped summations. A positive score is considered good,
otherwise it is considered bad.

There are only two good behaviors in the \gs\ \score: staying in the
mesh for a long time, and delivering messages on a subscribed topic
within some threshold time window.  The \score\ defines
five possible bad behaviors: delivering messages on a subscribed topic
at an insufficiently high rate, failing to quickly deliver a requested message
on a subscribed topic, sending an invalid message
(\eg, one that does not type-check), being co-located on the same IP
address as another peer, or trying to re-graft during the backoff
period after being pruned.  There is also an application-specific
\ind\ that can be positive or negative, and which is totally controlled
by the application.  This allows the application to reward or penalize
behaviors that it considers to be good or bad, respectively.

We denote the per-topic \inds\ with the shorthand $P_j(t)$, where~$j$ identifies the \ind\ and~$t$ is the topic.
This is the notation used by the \gs\ developers and should not be confused
with our lower-case notation for peers.
To be clear, when we discuss how the peer~$p$ uses the per-topic
\ind~$P_j(t)$ while scoring the peer~$q$, we are really referring to
$p.P_{j,q}(t)$, an \ind\ that is local to the scoring peer~$p$ and indexed by 
the \ind\ name~$j$,
    peer being scored~$q$, and
    topic~$t$.
Likewise, when we discuss how the peer~$p$ uses the global \ind~$P_k$
while scoring the peer~$q$, we are really referring to $p.P_{k,q}$, an 
\ind\ that is local to~$p$ and indexed by the \ind~name~$k$ and peer being scored~$q$.
For each topic~$t$,
    there are five topic-specific \inds: $P_1(t), P_2(t), P_3(t), P_{3b}(t),$ and~$P_4(t)$.  
The global \inds\ are $P_5, P_6,$ and~$P_7$.
All indicators are weighted in the \score\ with corresponding weights.

\noindent\textbf{Per-Topic \Inds.} We describe them below.

$P_1(t)$: \emph{Time in Mesh}. It is the amount of time quanta a peer
has continuously been a member of~$p.M(t)$ capped by
a small positive constant time in mesh cap that is
configured topic-by-topic by the application.
$P_1(t)$ is multiplied with a small positive
 constant configurable topical weight~$w_1(t)$.

 $P_2(t)$: \emph{First Message Deliveries}. The number of messages on
 the topic~$t$ for which the peer was one of our first deliverers (as
 measured by a constant, topic-specific, application-configured
 temporal threshold), multiplied by
 the topic-specific first message deliveries decay, 
 rounded down to zero if it
 falls below \texttt{DecayToZero}, and capped above by the positive
 corresponding first message deliveries cap.
$P_2(t)$ is multiplied with a positive
weight~$w_2(t)$. %

$P_3(t)$: \emph{Mesh Message Delivery Rate}. Let 
the mesh message deliveries on a topic~$t$ be
the number of messages delivered to~$p$ by~$q$ on~$t$,
multiplied at each time-step by a topic-specific
mesh message deliveries decay, and rounded to~0 if it
falls below \texttt{DecayToZero}. If 
the deliveries exceed a topical mesh message deliveries
threshold,
or if $P_1(t)$ does not exceed the topical mesh message deliveries activation,
then~$P_3(t)$ is set to~$0$.
Else,~$P_3(t)$ is the difference squared. 
$P_3(t)$ is multiplied by a weight~$w_3(t)<0$. %

$P_{3b}(t)$: \emph{Mesh Message Delivery Failures}.
Counts a random subset of the message delivery failures the scoring peer~$p$ 
observed from the peer~$q$ on the topic~$t$, multiplied at each time-step by
the topical mesh failure penalty decay, and rounded down to zero if it
falls below \texttt{DecayToZero}. A failure
occurs 
when~$q$ declares that it has a message~$x$, the scoring peer~$p$ responds
by requesting the message~$x$, and then 
$q$~fails to respond with $x$~in a timely
fashion. 
Whenever~$p$ prunes the~$q$, $p$
increments~$P_{3b}(t)$ by~$P_3(t)$. Ideally, this punishes pruned
peers so that they cannot quickly re-graft. 
$P_{3b}(t)$ is multiplied with a negative weight~$w_{3b}(t)$. %

$P_4(t)$: \emph{Invalid Messages}. The number of invalid
messages delivered to the scoring peer~$p$ by the
peer~$q$ on the topic~$t$, multiplied at each time-step by a 
topical invalid message deliveries decay and rounded to~0 if
it falls below~\texttt{DecayToZero}. 
Messages that does not type-check or that are marked invalid by the application are considered invalid.
$P_4(t)$
is multiplied with a negative weight~$w_4(t)$.

\noindent\textbf{Global \Inds.} We describe them below.

$P_5$: \emph{Application-Specific Score}. This is the score
  component assigned to the peer by the application itself. It is a
  real value that is multiplied in the \score\ by 
  a positive weight~$w_5$, so that the
  application can, e.g., signal misbehavior with a negative score, or gate peers before an application-specific handshake is completed.

$P_6$: \emph{IP Colocation Factor}. Let 
IP colocation factor refer to the number
of neighbors of~$p$ using the same IP address as the peer~$q$. If 
the IP colocation factor is not more than 
the IP colocation factor threshold, then $P_6$ is set to~0.
Else,~$P_6$ is set to the square of the difference. In the score
function,~$P_6$ is multiplied with a negative weight~$w_6$. 
This \ind~can be used to detect Sybils \emph{iff} the Sybils are IP co-located.

$P_7$: \emph{Behavioral Penalty}. Let 
the behavioral penalty be initialized at~0, incremented by the scoring peer~$p$ whenever~$q$ tries to re-graft 
less than \texttt{PruneBackoff} time after being pruned
or has a mesh message delivery failure, rounded down to 0 if it falls below \texttt{DecayToZero}, and multiplied by 
the behavior penalty decay at each \hbm\ event.
Let \texttt{excess} equal
the behavior penalty minus the behavior penalty threshold.  
Then $P_7=\texttt{excess}^2$ if 
the penalty exceeds the threshold, else 0.  In the \score, $P_7$ is multiplied with a negative weight~$w_7$.

\noindent\textbf{Configuring the \Score.}
The \gs\ specification 
    states that $w_1(t)$ should be a ``small positive'';
    $w_2(t)$ and $w_5$ should be ``positive'';
    $w_3(t), w_{3b}(t), w_4(t), w_6,$ and $w_7$ should be ``negative'';
    \texttt{DecayToZero}  should be ``close to 0.0'',
    the decay parameters should all be ``in (0.0, 1.0)'',
    time in mesh caps should be a ``small positive value'',
    first message deliveries caps should be 
        at least 
        the corresponding
        mesh message deliveries thresholds;
    IP colocation factor should be ``at least 1'',
    and
    mesh message deliveries thresholds should be ``positive'' and
    depend on the ``expected message rate for topic.''
This dependency is 
unexplained.
Guidance is likewise not provided for the topic weight tw$(t)$, nor for
topical mesh message deliveries activations~\cite{gs1.1}.

The \score\ requires additional peer-specific constants
configurable by the application.
The first is a non-negative constant \texttt{TopicCap},
used to define the function 
\(\text{TC}(x)=\min(x, \texttt{TopicCap})\) \emph{if} 
\(\texttt{TopicCap} \neq  0\) \emph{else} x.
This function limits 
the contribution of topic-specific behaviors
to the score.
Second, for each topic $t \in p.T$,
the \score\ requires a positive constant $\text{tw}(t)$ called the 
\emph{topic weight} of~$t$ controlling the relative influence of
 topic-specific behaviors to the score.  
The specification does not advise on how to configure the \texttt{TopicCap}
or topic weights.

 Note, the
 \gs~specification does not require peers to
 configure their \score s the same way.  In the case studies we
 considered (\fc\ and \Eth), nodes use identical configurations.
We enumerate the \score\ configuration variables in Table~\ref{tab:defaultParams} in the Appendix.

\noindent{\bf The \Score.}
Recall that $p.T$ denotes the set of all topics the peer~$p$ knows about,
    including those it does not subscribe to.
The \gs\ specification~\cite{gs1.1}
    defines 
    the score computed by a peer~$p$ 
    for a peer~$q$ as follows.
\[
 \mathit{score}(q)
 = \text{TC}\big(\sum_{t \in p.T} \text{tw}(t) (
 \underset{i \in \{1,2,3,3b,4\}}{\sum w_i(t) P_i(t)} ) \big) 
 + \sum_{i=5}^7 w_i P_i
\]

\subsection{Attack Mitigation Using the \Score}\label{subsec:defenses}

\gs~leverages heuristic defense mechanisms
based on
two caches, \emph{mcache} and \emph{seen}.  The \emph{mcache} stores full messages and their identifiers, enabling lazy pull.  To avoid memory overflow, it is partitioned into lists called \emph{history windows}.  Periodically, a new history window containing the most recently sent or received new messages is pushed to the cache, and if the cache size exceeds a parameter \texttt{McacheLen}, then the oldest history window is deleted.  The \emph{seen} cache is a timed cache, but only tracks message identifiers and is used to avoid infinite forwarding loops.
The defense mechanisms and their caches are tuned by a set of parameters, 
detailed in the Appendix in Table~\ref{tab:defaultParams}.

\noindent~$\bullet$~{\em Pruning.}
This mechanism is \emph{controlled mesh (and fanout) maintenance}.
Peers whose scores fall below zero are pruned from the mesh and fanout 
at every heartbeat maintenance event (by default, every second).
  
\noindent~$\bullet$~{\em Opportunistic Grafting.}
 The goal of this mechanism is to 
 add peers
  who behave properly (and thus accumulate positive score) to the mesh- and fanout-peer sets.
If the median score of fellow mesh peers
  is below the threshold \texttt{OpportunisticGraftThreshold}, 
  then above-median scoring neighbors 
  are opportunistically grafted.

\noindent~$\bullet$~{\em Backoff on Prune.}
This mechanism adds a backoff 
    period \texttt{PruneBackoff} after pruning during which the pruned peer is forbidden
    from re-grafting, ensuring that pruned
nodes cannot quickly rejoin.

\noindent~$\bullet$~{\em Flood Publishing.}
To limit the impact of attacks when a message is first sent,
\gs\ includes an optional \emph{flood publishing} feature, where each peer sends every
newly published message to all topic-subscribed neighbors whose scores exceed the positive \texttt{PublishThreshold}.
  This ensures that a new message is disseminated to properly behaving peers (who presumably have high scores)
        even when the network is saturated with malicious nodes.
    
\noindent~$\bullet$~{\em Adaptive Gossip Dissemination.} 
In \gs 's lazy pull mechanism, peers adaptively update the number of neighbors
to whom they emit topical gossip.
The feature is designed to achieve some benefits of flood
publishing without all the bandwidth cost, to combat
malicious nodes.

\subsection{Previous Attack Analysis of \gs}

The Protocol Labs ResNetLab and software audit firm Least Authority tested \gs\ against a list of specific pre-programmed attack scenarios (\eg, malicious peers saturate a network and simultaneously stop transmitting data) using a network emulator called \textsc{Testground}~\cite{testground}.   
In each simulation, the attacker goal was to degrade
network performance, \ie, to increase average dissemination time
and loss.
They used simplified configurations with only one topic,
and their configurations did not exactly match those currently used by \fc\ and \Eth,
as these values have since been updated (partially as a consequence of their findings).
They simulated 1,000 honest peers and 4,000 malicious nodes, allowing
    each malicious node to establish 100 connections, 
    and each honest peer to establish 20.
They also tested
    the \bc\ and \OldEth\ gossip protocols,
    as well as \gs\ without the defense mechanisms.
Their success criteria for the defense mechanisms
    were that messages were fully disseminated 
    in $<$ 6s for \fc\ or $<$ 12s for \Eth, and
    the loss rate was low.
    (They did not specify what they considered to be low.)
They found that all the attacks failed against \gs,
and the defense mechanisms made \gs\ more resilient than other tested protocols to attacks by malicious nodes~\cite{vyzovitis2020gossipsub}.
Separate from simulation testing, Least Authority 
also audited the \golang\ implementation and provided recommendations for improvement~\cite{leastAuth20}.

\textbf{\gs~vs. Flooding.}
Flooding is the only protocol that guarantees message delivery between two parties, as long as there is an
adversary-free path between them. However, flooding achieves this by sending data over all possible paths.
Another approach is to send only on $k$ paths, for some~$k$,  but it requires
disjoint paths and assumes the attacker does not control more than $k-1$ paths, which is not always realistic for real networks. 
\gs\ was proposed to provide similar security without requiring sending data on all paths, or requiring disjoint paths,
and its authors showed experimentally that under certain scenarios it does prevent some attacks. The goal
of our work is to understand formally what security is actually provided by the score function.

One of the main limitations of flooding is that it incurs high communication overhead because it redundantly
sends messages over the network, and this overhead is incurred
regardless of the presence of misbehaving peers. \gs\ was designed to address this limitation.  It incurs less network load than flooding because it sends messages only to peers that need them.  Specifically, it sends meta-data to a limited number of peers
(as opposed to flooding, in which full messages are sent to all peers).
Then peers who recieve this meta-data can request the specific messages they need.
The \score\ has no additional communication cost because it is computed based on messages that are normally generated by the system and the scores are never exchanged. Thus, in the case when there are no misbehaving nodes, \gs~has small communication overhead. In the case when there are misbehaving nodes, 
 the defense mechanisms built into \gs\ based on the score function can flag certain kinds of malicious peers, e.g., they can detect when a large number of peers share a single IP.  The cost of defending against these attacks is adjusting the mesh by removing the misbehaving peers. 


\section{\acls~\gs~Model}
\label{sec:model}
We used \acls\ to model and reason about \gs. We briefly discuss our
code in the Appendix.  Our model captures every aspect of \gs\ given
in the prose specification\ \cite{gs1.0,gs1.1}: control messages, lazy
pull, the internal peer state, meshes and fanouts, the \score, the
defense mechanisms, etc., including every detail in
Section~\ref{sec:gs}, and other prose specification
details that we omitted for
readability and space. Despite the fact that our model is fully
faithful to the written specification of \gs, and is itself an
executable program, it is not a network library; it cannot be used in
place of the \golang\ implementation; and it is not intended to be
used as such. It is simply an in-memory model of \gs, which also
happens to be a mathematical object that we can reason about using
\acls.

Recall that a core feature of \gs\ is its use of heuristic defense
mechanisms to promote well-behaved peers and demote poorly-behaved
ones, \eg, by grafting or flood-publishing to high-scoring peers while
pruning peers with negative scores. We state this feature as the
\emph{fundamental property} of the defense mechanisms. We formalize
four novel correctness properties for the \score\ that are necessary
for this fundamental property to hold, focusing on the most
  general properties of the \score. Note, these properties do not
comprehensively cover all gossip protocols.  The properties for a
gossip protocol depend on the application, \eg, one might prioritize
dissemination speed, another reliability.   We
solicited security properties from the \gs\ developers and the
Ethereum Foundation; both endorsed our properties but did not provide
more. We test our four properties using the counterexample generation
facility provided by \acls, and find counterexamples to two of them,
while \acls\ semi-automatically proved the third and fourth. Finally, we
synthesize traces that lead to such counterexamples, and show that
they also violate the fundamental property of the defense
mechanisms. The sequences of actions taken by adversary peers in these
traces constitute attacks against \gs.  In these attacks, adversaries
(attackers) misbehave by not forwarding data, thereby slowing down the
entire network while avoiding getting
  pruned.

\textbf{Modeling Assumptions} We make the following assumptions:
  (1) the message payload can be abstracted by a record consisting of
  a message-id and a message, both represented by natural numbers
  (since our properties and attacks do not depend on message
    content); (2) the transport protocols by which \gs\ sends and
  receives messages can be represented using a partial ordering on
  message send and receive events; (3) message transmissions are not
  lost, duplicated, or corrupted, but can be reordered; (4)
  peer-discovery took place prior to model instantiation; (5)
  connection establishment is abstracted with \texttt{CONNECT}
  events; and (6) we assume the existence of an oracle for making
  non-deterministic choices and we use this to formally reason about
  different plausible choices a peer might make.

\subsection{Validating Our Model}
Our model allows us to execute and reason about any component of the
peer logic in isolation, the entire program for a peer, or even an
entire network of peers.  We developed our model in consultation with
the \gs\ authors, who asked us to study the \score\ in their protocol.
We validated our model in multiple ways.  We implemented all the tests
from their \golang\ code as tests or theorems in our model;
instrumented the \golang\ code to print traces, which we type-checked
with our model; and generated counterexamples in our model, which we
translated into (passing) \golang\ unit-tests.  These
  conformance checks awarded us high confidence that our model closely
  matches the \gs\ protocol described in the specification document,
  as well as its Golang implementation.  Note however that our
  approach did not involve instrumenting the model and implementation
  to directly communicate with one another, e.g., as was done
  in~\cite{bhargavan2021depth}.  Through these exercises, we found
ambiguities in the prose and places where the code and prose
disagreed. We reported errors to the developers and followed their
advice to resolve ambiguities.
  
\textbf{Discrepancies.}  In the specification, but not the
implementation, the activation window is used when calculating $P_3$
and $P_{3b}$.  In the implementation, $P_3$ is updated periodically,
but in the specification it is updated only when the peer is pruned.
Components of the \score\ can be disabled in the implementation but
not the specification.
We allowed disabling in our model because \fc\
uses this feature, but otherwise we followed the English specification.

Our \acls\ model is fully verified and contains 6,768 lines of code,
203 definitions, and 177 explicit theorems and properties.  Every
function definition involves proofs that are not included in the
explicit theorem total, \eg, of termination, that the input contracts
imply the output contracts,
etc. Most model development effort was devoted to
translating the prose specification into a mathematical form,
comparing it to the \golang\ implementation, and translating tests from
\golang\ into \acls. Once we had the model, it was fairly
straightforward to write properties to test for
counterexamples. Developing and admitting
all our functions, with full termination and contract proofs required effort comparable to developing and
unit-testing the corresponding functions in a traditional
implementation.

\subsection{Model State and Transitions}
We define the state of a peer using the \ps\ type and of multiple
peers using the \group\ type.
\begin{lstlisting}[language=Lisp]
(defdata peer-state
  (record (nts . nbr-topic-state)
	  (mst . msgs-state)
	  (nbr-tctrs . pt-tctrs-map)
          (nbr-gctrs . p-gctrs-map)
	  (nbr-scores . peer-rational-map)))

(defdata group (map peer-id peer-state))
\end{lstlisting}
Each \ps\ for a peer\ $p$ is a record consisting of five components:
(1) \texttt{nts} of type \nts: a state containing $p$'s neighbors'
subscriptions, $p.M$, $p.F$, and a map storing $p$'s last publication
time in each topic (used for fanout maintenance); (2) \texttt{mst} of
type \mst: a state containing a cache of full messages received, a map
from recently seen message ids to their age (updated at every heartbeat maintenance event), history windows and the completion status and
count of both sent and received requests for messages; (3)
\texttt{nbr-tctrs} of type \ptt: a total map from each pair of
neighbor $q$ and $t \in p.T$ to \tctrs; (4) \texttt{nbr-gctrs} of type
\texttt{p-gctrs-map}: a total map from each neighbor $q$ to a list of
global counters $\langle P_i \ | \ i \in \{ 5,6,7 \} \rangle$; and (5) \nbrscores\
of type \texttt{peer-rational-map}: a total map from neighbors of $p$
to their cached scores, which gets updated at every local \hbm. All
scores and
counters default to 0.  Peers are identified using unique \p s, and a
\group\ is simply a finite map (an association list) from \p\ to \ps.

Notice that all \ps s\ in a \group\ are simultaneous: a
\group\ captures an instantaneous snapshot of the \gs\ network.
However, the peers themselves do not have access to a global
clock. They are only aware of the partial ordering of events they can
locally infer. A peer does not have access to any other \ps\ besides
its own, nor to any data that is not locally observable.

We define \gs\ network events using a type called\ \evnt. An
\evnt\ occurs when a peer sends or receives a control message, joins
or leaves a topic, goes through a local \hbm\ event, forwards a message
from the application layer, or establishes a connection with another
peer. Events where messages are sent or received
carry the identity of two peers: sender and receiver.  Every event
carries the identity of the peer who triggered it, as its first
element.  The \evnt\ type is described using BNF below,
where \texttt{pid} is a peer identifier, 
\texttt{vrb} is \texttt{SND} or \texttt{RCV},
\texttt{top} is any of the topics in the application
  running on \gs, and \texttt{msg} is any payload including
  control- or full-messages. 
  
\begin{verbatim}
event ::= pid vrb pid msg | pid JOIN top
        | pid CONNECT top | pid HBM top
        | pid APP top msg | pid LEAVE pid
\end{verbatim}

  Every application that runs on top of \gs\ tunes weights and
  parameters in order to define the \score\, such as tw$(s)$,
  $w_3(s)$, the mesh message deliveries decay on~$s$, etc., as
  detailed in Section~\ref{subsec:score}.  We store topic-specific
  weights and parameters in a map from topics to corresponding weights
  and parameters, which we call \twp.  Note that \twp\ is a constant
  specific to the application instance we are simulating. The \ps\
  transition function is called \texttt{ps-trx} (illustrated in
  Figure~\ref{fig:peer-state-fsm}).  It takes as input a \ps\ called
  \texttt{ps}, a \twp, and an event called \texttt{evnt}.  We assume
  an oracle for making non-deterministic decisions in the model.  In
  simulation runs the oracle can be replaced by a pseudo-random number
  generator, for convenience.  \texttt{ps-trx} outputs the peer's new
  \ps, as well as the \evnt s it emits during the transition.

  The \group\ transition function is called \texttt{gs-trx} and takes
  as input a \group\ called \texttt{gp}, a \twp, and a work-list of
  \evnt s initialized with \texttt{init-evnts}.  \texttt{gs-trx}
  assumes the same oracle as \texttt{ps-trx}.  The peer who triggered
  the first \evnt\ in a work-list of \evnt s transitions on that
  \evnt\ using the \ps\ transition function. A new \group\ is then
  generated where the peer's old \ps\ is replaced with its new one.
  Along the way, it also computes any emitted \evnt s which are
  appended at the end of the \evnt s work list.  The function is
  illustrated in Figure~\ref{fig:group-fsm}.

\subsection{\gs\ \Score\ Properties}\label{subsec:props}
The most important features of \gs\ are lower bandwidth consumption
via lazy pull, and security against malicious peers via heuristic
defense mechanisms.  The fundamental idea of the heuristic defense
mechanisms is that honest peers can be distinguished from malicious
ones based on their observable behaviors, and thus, the overall
network can be made more secure and performant if every honest peer
promotes their well-behaving neighbors and demotes poorly-behaved
ones.  We formalize this requirement as follows, where poor and good
behavior are defined by the bad and good
behavior counters given in
Section~\ref{sec:gs}.
\begin{description}
\item \textbf{Fundamental Property of the Defense Mechanisms.}
  \emph{Peers who behave poorly will be demoted by their neighbors.
    Peers who behave better-than-average will be promoted by their
    neighbors.  Promotion/demotion is entirely based on peer
    behavior.}
\end{description} 
Studying this fundamental property directly is difficult due to the
massive search-space of possible attack vectors.  Hence, we decompose
the problem by proposing four novel security properties for the
\score\ without which the \emph{fundamental property} cannot possibly
hold.  We choose these properties such that a reasonable software
developer might infer that they are true about \gs, based on textual
descriptions of the protocol by the \gs~developers.  We encode these
properties in \acls~as predicates over data types defined above.  The
properties are defined in an app-specific manner, \ie, each property
is parameterized by a fixed app-specific \twp.  The fundamental
  property is written for human consumption, and is informal.  In contrast, our four formal
  properties unambiguously define the fundamental one in a
  way that is amenable to formal
  verification.

Importantly, all four properties are independent of the number of
peers, percentage of malicious peers, or network topology.  They
depend only on the topic, app-specific parameters, and performance
counters of the peer being scored.

The \gs\
developers write~\cite{vyzovitis2020gossipsub}: \emph{The \score\ is used
as a performance monitoring mechanism to identify and remove poorly
performing or misbehaving nodes from the mesh.}  Since meshes are
topic specific, we naturally ask, does the \score\
identify poorly performing nodes in each topic?  Peers can
subscribe to, and forward messages over several topics, hence a peer can be a member of several meshes.  As peers that
accumulate a non-positive score get
pruned, we claim that continuously achieving a non-positive score in
a topic should eventually result in a non-positive overall score,
leading the peer to be pruned.  If this is not true, then
neither is the fundamental property, as one of the defense
mechanisms is that poorly-behaved peers get opportunistically pruned.
For example, in Fig.~\ref{fig:mesh-fanout-diagram}, if B
throttles deliveries in topic \blue\ to A, then we ask if A
will assign a negative score to B and thus prune it during maintenance.
We formalize this liveness property below.
\begin{property}\label{p1}
  \emph{If a peer's score relating to its performance in any topic is
    continuously non-positive, then the peer's overall score should
    eventually be non-positive:}
  \[\begin{aligned}
  \forall q,t:: & \langle
  \glb(\mathit{score(q)}\textit{ for topic }t \leq 0) \Rightarrow \\& \,\,\,\fin(\mathit{score(q)} \leq 0) \rangle 
  \end{aligned}\]
\[
\text{where score$(q)$ for topic $t$ is}
  \text{tw}(t) (\underset{i \in \{1,2,3,3b,4\}}{\sum w_i(t) P_i(t)}).
\]
\end{property}   
The \gs~developers write that peers ``that misbehave are penalized
with negative score.''~\cite{vyzovitis2020gossipsub} This feature is
important, because the opportunistic grafting and mesh and fanout
maintenance defense mechanisms of \gs\ assure that over time a peer
disconnects from neighbors who have negative or below-average scores
and connects to those who have positive scores.  So, if a peer could
misbehave in a specific topic, without getting penalized with a
negative score, then these defense mechanisms would be ineffective and
the fundamental property would be violated.

The next three are safety properties. We identify the
following as bad-performance metrics indicating misbehavior: deficit
in mesh message deliveries (DMMD), invalid message deliveries, and bad
behaviors; where: DMMD is the maximum of 0 or the mesh message
deliveries threshold minus the mesh message deliveries.  Note, these
metrics are used in the \score.  When discussing peers $q, q'$ in the
properties below, we use $P_j, P_i(t)$ to denote \inds\ of $q$ and
$P_j', P_i'(t)$ to denote \inds\ of $q'$.
\begin{property}
\label{p2}
\emph{Increasing bad-performance counters should decrease overall
  score.  Formally, if $P_i'(t)$ differs from $P_i(t)$ only due to an
  increase in DMMD, invalid message deliveries, or bad behaviors for
  peer $q$ in topic $t$, then:}
\[\forall q,t::\langle
  (\mathit{score(q)}\textit{ for topic }t)  > (\mathit{score(q')}\textit{ for topic }t)  \rangle \]
\end{property}
A simplified \acls\ definition of the contraposition to this property in
context of \Eth\ is shown in Figure~\ref{lst::p1}, where
\texttt{*eth-twp*} is a \twp\ specific to \Eth.

The \gs~developers write that the role of $P_1(t)$ in the \score~is to
``boost peers already in the mesh'', and the role of $P_2(t)$ is to
``reward peers who act fast on relaying messages.''  The
app-specific component~$P_5$ ``has an arbitrary real value, so
that the application can signal misbehavior with a negative score''
or good behavior with a positive score~\cite{vyzovitis2020gossipsub}.
We define good-performance counters (that measure good behavior) as
mesh time, first message deliveries, and mesh message deliveries, and claim that 
increasing one of these counters should boost the overall
score, implying the following analogue to Property~\ref{p2}:

\begin{property}
\label{p3}
\emph{Increasing good-performance counters will not decrease score for
  a mesh peer that has been in the mesh for a sufficiently long
  time. Formally, if $P_i'(t)$ differs from $P_i(t)$ only due to
  increase in mesh time, first message deliveries, or mesh message
  deliveries for peer $q$ in topic $t$, and the mesh time is more than
  the \texttt{activationWindow} parameter, then:}
\[\forall q,t::\langle
  (\mathit{score(q)}\text{ for topic }t)  \leq (\mathit{score(q')}\text{ for topic }t)  \rangle \]
\end{property}

In \gs, ``all nodes start equal and build their profile based on their
behavior''~\cite{vyzovitis2020gossipsub}.  Concretely, the \score\ is
referentially transparent: a peer's score is a function of its
behavior alone.  Hence the \score\ is intrinsically unbiased, \ie, if
two peers behave identically, then they will achieve identical scores.
If this property were not true, then the controlled mesh (and fanout)
maintenance and opportunistic grafting defense mechanisms would behave
unfairly with respect to peer behavior/misbehavior, potentially
violating the fundamental property by making promotion and demotion
decisions not based on the good and bad behavior counters.  We
formalize this in Property~\ref{p4}.

\begin{property}
\label{p4}
\emph{If two peers subscribe to the same topics $ \in S$, and 
achieve identical per-topic params $P_1(t)$, $P_2(t)$, $P_3(t)$, $P_{3b}(t)$, $P_4(t)$, $\forall$ $t \in T$,
and identical global params $P_5$, $P_6$, $P_7$;
then they achieve identical scores.}
\end{property}

\subsection{Finding Counterexamples}
\label{subsec:counterexample}
Our model can be used not only to reason about and simulate a \gs\
network, but also, to automatically disprove invalid properties by computing
concrete counterexamples. 
We generate counterexamples with our model using the cgen
library built into \acls, which uses type enumerators,
synergistically combined with theorem proving techniques, to generate
values for variables within a property such that the hypotheses of the
property hold but the implication does not. However, the sample space
for counterexamples is huge. Hence, we need to define our own
\emph{custom} enumerators for the types of each of these variables,
using our intuition about these variables and their types such that
hypotheses in our properties are almost always satisfied thereby
increasing the chances of discovering values that actually violate the
property. A custom enumerator for a type $\tau$ is a function from
naturals to type $\tau$. When necessary, we craft custom
enumerators to quickly find interesting counterexamples. These
counterexamples are snapshots of a \gs~network where the property
being tested is violated. Importantly, the snapshot might not be a
reachable state of the network. We interpret these counterexamples as
attacker specifications: an attacker can attempt to violate the tested
property by guiding the network into satisfying one of these
counterexample specifications.  Put differently, a counterexample is
just a bad network state an attacker would like to
achieve, whereas an attack is a sequence of actions performed by one
or more attackers that guides an initial network state to a
counterexample one.  In such attacks, each attacker violates the
fundamental property of the defense mechanisms, misbehaving (\eg, by
not forwarding data, or by sending invalid messages) while being
promoted or without being demoted by its neighbors.

\subsection{Generating Attacks}
We generate attacks by generating counterexamples to security
  properties under reasonable security assumptions.  First, we make
  all of the assumptions listed in Section~\ref{subsec:gs-overview}.
  Second, we assume only a minority of peers in the network are
  the attackers, and their goal is to throttle or block the
  dissemination of messages from honest peers, without being detected
  by the honest peers. The attack generation process goes as follows. We begin with a
  specific \group\ representing the initial state of the network,
  satisfying our assumptions. Given a counterexample
  \group\ violating a \twp-parametrized property, we ask if the
  attacker(s) could, under our assumptions,
  guide the initial \group\ to the counterexample \group\ (or a
  similar one).  The state-space of \trc s is
too large to be explored by the counterexample generation facility
alone.  Instead we generate a
  list of events (each of type \evnt) to lead the initial \group\ to a
  similar counterexample \group, based on the assumption that all
  peers in the group behave honestly, except the attacker peer.

Note that our four properties are defined entirely with respect to the
\score.  They do not take into account the network topology, the
number of malicious versus honest peers in the network or their
placement, the network dynamics (throughput, churn, etc.), or other
variables whatsoever, except for those used by the \score.
Thus, when we prove that a property holds for a given \twp,
our proof shows that the property holds for \emph{all possible \gs\
  networks} configured by that \twp, regardless of their topology,
placement and number of malicious peers, etc.  Conversely, if we prove
that the property does not hold for a given \twp, then we know we can
attack \emph{every single \gs\ network} configured with that \twp,
although the attacks themselves need to be generated using the network
state (e.g. topology, number and placement of malicious peers).

\subsection{Evaluating New Application Configurations}
\label{subsec:configurations}
One advantage of our model is that it allows developers of new
applications using \gs\ to check if their configuration satisfies our
security properties. The developer needs to formalize the
configuration as a \twp, instantiate the properties using that \twp,
and then pass the resulting model file into \acls.  \acls\ may prove the properties automatically,
outputting \texttt{qed}; it may output counterexamples, as it does
for \Eth\ (in which case the configuration should be debugged using
the counterexamples); or it may fail to do either.  In the last
scenario, the developer may either tweak the enumerators with which
\acls\ generates its counterexamples, or guide the prover using
supplemental lemmas, until the properties are disproven or proven.  We
exemplify how to tweak the enumerators in \emph{scoring-eth2.lisp},
and how to guide the prover using lemmas in \emph{scoring.lisp}. Both
are part of our publicly available materials.


\section{Experiments}
\label{sec:exp}

In their emulation testing, the \gs\ developers checked if the defense
mechanisms improved the resiliency of \gs\ to specific attacks from
misbehaving peers against network performance.  We ask a more
fundamental question: does the \score\ upon which the defense
mechanisms rely actually measure what it is intended to measure?  If
the answer is \emph{no}, then there might exist covert attack
strategies that are undetectable using the \score\ with certain \gs\
configurations.  This question is articulated via our four formal
properties and evaluated on two concrete case studies: \Eth, and \fc.

\subsection{Methodology}
As described in Section~\ref{sec:model}, the \acls\ model is infinite
state and faithful to the specification. Any properties we prove hold
for all of the infinite instances of the model, with any peers,
topology, set of topics, history of events, etc. If a property fails,
then there exists a counterexample, but typically infinitely
many. When generating counterexamples, we prefer to use minimal
networks to ease readability and improve generation efficiency.

Our properties do not depend on the percentage of
malicious peers or network topology or size. Hence,  to
find vulnerabilities and eventually exploit them, we can instantiate a
small \gs\ \group\ in our model, using the corresponding
app-specific \twp.  The simple \group\ consists of two honest
peers and one attacker, all fully mesh connected on every topic in the
\twp, and allows us to explore the event-space very
quickly. Our app-specific \twp s are shown in
Tables~\ref{tab:fctwp} and~\ref{tab:ethtwp} and are adapted from
Github open-source implementations of  \Eth, and \fc.  We
use \acls\ to try and generate counterexamples with each case study
\group, for each property.

If we find counterexamples, we attempt to generate corresponding
  attacks. These attacks are not like those considered in the
  emulations done by the \gs\ developers. Rather, they describe how a
  peer can violate one of the properties, \eg, by misbehaving without
  decreasing its score.  Such attacks in a small \gs\ \group\ can be
  viewed as the building blocks for crafting stealthier or more complex attacks, e.g. eclipsing a peer in a targeted topic.

\subsection{\fc\ Score Function Properties Evaluation}
We prove that the \fc\ \twp\ (Table~\ref{tab:fctwp}) satisfies
  all four properties.  Unfortunately, the \fc\ \twp\ satisfies the
  properties by violating the \gs\ specification, in that it uses
  illegal 0-valued weights and thresholds, sacrificing its ability to
  penalize peers who under-deliver. Since the \fc\ \twp\ disables the
  app-specific score component, the application also cannot signal
  level misbehavior.  Hence the \gs\ layer of \fc\ is (in isolation)
  less resilient to attacks.  (The \fc\ developers inform us that for
  this reason, \fc\ relies on app-level defenses.)

\subsection{\Eth\ Score Function Properties Evaluation}

\noindent   
We auto-generate counterexamples to
Props.~\ref{p1} and \ref{p2}.
\\
\noindent \textbf{Prop.~\ref{p1}.}  
\Eth\ violates Prop.~\ref{p1} because it has multiple topics and a
peer can offset a negative score in one of the (dozens) of subnet
aggregator topics by a positive score in all other topics combined,
resulting in a positive overall score.  A counterexample is shown in
Tables~\ref{tab:eth-ctrex1} and~\ref{tab:eth-ctrex2}.  In
Table~\ref{tab:eth-ctrex1}, an \Eth\ peer under-performs in topics
\texttt{AGG} and \texttt{SUB2}, \ie, its topic specific counters FMD
and MMD are less than the threshold required for message
delivery. However, its performance is nominal in the rest of the
topics.  Table~\ref{tab:eth-ctrex2} shows \inds\ computed from
topic-counters in Table~\ref{tab:eth-ctrex1}. Note non-zero $P_3$ and
$P_{3b}$ in \texttt{AGG} and \texttt{SUB2}, leading these topics to
negatively impact the score.

\begin{table}
  \begin{center}
    \begin{tabular}{|l|l|l|l|l|l|}
      \hline
      Topic  & MT & FMD & MMD & IMD & MFP \\ \hline
      \texttt{Blocks} & 147  & 194 & 200   & 0   &  0     \\ \hline
      \texttt{Agg}    & 42   & 0    & 1    &0    &  0  \\ \hline
      \texttt{Sub1}   & 141  & 188 & 194     & 0    & 0     \\ \hline
      \texttt{Sub2}   & 42    & 0   & 1   & 0  & 0   \\ \hline
      \texttt{Sub3}   & 135  & 182  & 188   & 0    &  0    \\ \hline
    \end{tabular}
  \end{center}
  \caption{\Eth\ peer \tctrs\ violating
    Prop.~\ref{p1}}
  \vspace*{-2em}
  \label{tab:eth-ctrex1}
\end{table}

\begin{table}
  \begin{center}
    \begin{tabular}{|l|p{0.4cm}|p{0.3cm}|p{0.3cm}|p{0.3cm}|p{0.3cm}|p{0.3cm}|p{0.3cm}|p{0.3cm}|l|}
      \hline
      Topic  & $P_{1}$ & $P_{2}$ & $P_{3}$ & $P_{3b}$ & $P_{4}$ & $P_{5}$ & $P_{6}$ &
                                                                    $P_{7}$
      & {score} \\ \hline
      \texttt{Blocks} & 147  & 23   & 0     & 0        & 0       & 0       & 0       & 0       & 22.21       \\ \hline
      \texttt{Agg}     & 42       & 0   & 81  &81        & 0       & 0       & 0       & 0       & -4.5       \\ \hline
      \texttt{Sub1}   & 14.1   & 24   & 0   & 0        & 0       & 0       & 0       & 0       & 7.80       \\ \hline
      \texttt{Sub2}   & 4.2       & 0   & 1   & 1        & 0       & 0       & 0       & 0       & -25      \\ \hline
      \texttt{Sub3}   & 13.5    & 24   & 0  & 0    & 0       & 0 & 0       & 0       & 7.78      \\ \hline
    \end{tabular}
  \end{center}
  \caption{\Eth\ peer score components violating
    Prop.~\ref{p1}.  Total score = 8.29.}
  \label{tab:eth-ctrex2}
  \vspace*{-2em}
\end{table}
\noindent \textbf{Prop.~\ref{p2}.}
The \Eth\ \twp\ \texttt{TopicCap}$=37.72$. However, the sum of contributions to score from
each topic can be well above this limit, violating Prop.~\ref{p2}.  
A counterexample is shown in Table~\ref{tab:eth-ctrex3}, where, even after
perturbations to FMD and MMD give rise to lower values in FMD' and
MMD', the overall score remains 32.72.
In contrast, \fc\ does not use a \texttt{TopicCap}, and thus satisfies this property.

\noindent \textbf{Prop.~\ref{p3}.} We prove that this property holds for all valid configurations because the positive contributors to the score within a topic are monotonic.

\noindent \textbf{Prop.~\ref{p4}.} This property is proved
simply by getting the \score\ admitted in \acl. Since \acls\ is a
functional language, admitted functions will always have the same
output for same inputs, thus validating this property for any possible
application running on \gs.

\begin{table}
  \begin{center}
    \begin{tabular}{|p{1cm}|p{0.3cm}|p{0.3cm}|p{0.3cm}|p{0.3cm}|p{0.3cm}|p{0.3cm}|p{0.3cm}|r|r|}
      \hline
      Topic  & \rotatebox{90}{MT} & \rotatebox{90}{FMD} & \rotatebox{90}{FMD'} & \rotatebox{90}{MMD} & \rotatebox{90}{MMD'} & \rotatebox{90}{IMD} & \rotatebox{90}{MFP} & \rotatebox{90}{Score} & \rotatebox{90}{Score'} \\ \hline
      \texttt{Blocks} & 147  & 194 & 3   & 200 & 10  & 0 & 0 & 22.21 & 6.21 \\ \hline
      \texttt{Agg}    & 150  & 194 & 194 & 230 & 230 & 0 & 0 & 13.83 & 13.83 \\ \hline
      \texttt{Sub1}   & 141  & 188 & 188 & 194 & 194 & 0 & 0 & 7.80  & 7.80     \\ \hline
      \texttt{Sub2}   & 110  & 180 & 180 & 232 & 232 & 0 & 0 & 7.80  & 7.80  \\ \hline
      \texttt{Sub3}   & 135  & 182 & 182 & 188 & 188 & 0 & 0 & 7.78  & 7.78    \\ \hline
    \end{tabular}
  \end{center}
  \caption{Perturbations in \tctrs~values 
  for an \Eth~peer which violate
    Property 2.  Both totals = 32.72.}
  \label{tab:eth-ctrex3}
  \vspace*{-2em}
\end{table}

Based on the insights gleaned from studying counterexamples generated
by \acls\ for Props. 1-3, we manually crafted a pathological \twp\
(Table~\ref{tab:pathtwp}) with reduced penalties for low \mmd.

\subsection{Synthesizing Attacks for \Eth}
\noindent
\textbf{\Eth\ Topologies.} We synthesize attacks using the
real network topologies of the \Eth\ testnets Ropsten, Goerli, and
Rinkeby, as measured by Li et. al.~\cite{toposhot}.
Table~\ref{tab:eth-topos} shows basic characteristics of these
\Eth\ network topologies.

\begin{table}
\begin{center}
\begin{tabular}{|l|r|r|r|r|r|}
  \hline
  \multirow{2}{*}{Network} &
  \multirow{2}{*}{Nodes} &
  \multicolumn{3}{c|}{Degree} &
  \multirow{2}{*}{Diameter} \\
  \cline{3-5}
  && min & max & avg & \\
  \hline
  Ropsten  & 588   & 1  & 418   & 25.49  &  5 \\ \hline
  Goerli   & 1355  & 1  & 712   & 28.26  & 5  \\ \hline
  Rinkeby  & 446   & 1  & 191   & 68.96  & 6  \\ \hline
\end{tabular}
\end{center}
  \caption{\Eth\ Network Characteristics}
  \label{tab:eth-topos}
\end{table}

\noindent
\textbf{Attacks.}
We consider the following attacks:\\ 
\noindent~$\bullet$~Block/Throttle -- a single attacker
who shares a number of topics with a single victim
throttles/blocks the target topics without his score being decreased.
(A throttling attack limits communication within a topic, 
whereas a blocking blocks said communication entirely.)\\ 
\noindent~$\bullet$~Eclipse --  multiple attackers
surround the victim and block 
target topics for it. Note that 
traditional eclipse attacks where attackers will block 
{\em all topics} will be prevented by the score function. The property violation
that we found will allow instead for an attacker to block specific
topics without having his score decreased.\\
\noindent~$\bullet$~Partition -- multiple attackers target multiple victims by blocking
the target topics for each.\\
We abstract the essence of the vulnerability we discovered
in a gadget. Our attack gadgets can be applied to any network topology and allow an
attacker to block or throttle certain message transmissions, without
being discovered and without incurring any penalties. We show
that for all of these attacks the scores assigned to the attackers by
the victims stabilize; hence, by induction, they remain positive
forever.

\noindent
\textbf{Attack Gadget.} An attack gadget is a tuple
$\langle A, V, S \rangle$, where 
the attacker $A$ and victim $V$ are peers, $S$ is a set of subnet topics under attack,
and $A$ and $V$ are mesh neighbors over a set of topics that is a
superset of $S$. For each $i \in \mathds{N}$, we define \ag{i} to be the
set of attack gadgets where $|S|=i$. 
The attack gadget
allows $A$ to maintain an overall positive score while misbehaving
with respect to $S$, by behaving honestly with respect to the other
topics.  Therefore, if $T$ is the set of subnet topics in the given
\Eth\ network, an attack gadget $\langle A, V, S \rangle$ in \ag{i} is only
possible if $|T \setminus S|$ is large enough.  Using the \Eth\ \gs\
parameters, we can calculate the number of other topics $A$ and $V$
have to subscribe to as follows.
\begin{equation}
  \min \{t \in \mathds{N} \ | \ (7.2 + 3.2\frac{t}{T} > 24.7\frac{i}{T}) \wedge (t + i
  \leq T)\}
  \label{fmla}
\end{equation}
Eqn.~\ref{fmla} only has a solution for some values of $T$. 
To derive a corresponding formula for a different \gs\ application,
one has to use that application's parameters.

\noindent
\textbf{Experimental Setup.} 
 Given a topology and the number~$T$ of subnet topics, we
construct the corresponding model (of type \group), with topics
\texttt{BLOCKS} and \texttt{AGG}, in addition to $T$-many subnet topics, for a total of $n=T+2$
topics.  Every peer is a mesh member of every
topic. We then generate attacks and check that the attacks are successful, \ie, to check that the attackers
\emph{continuously} limit messages on the targeted topics without
\emph{ever} being penalized by the victims.

\noindent
\textbf{Throttle/Blocking Attacks.} We create the attack by
instantiating a
single attack gadget. For each network topology and number of attacked topics $i
\in \{1, 2, 3\}$, we generate a corresponding \acls\ model. In the model, we
determine the number of subnet topics~$T$ using
Eqn.~\ref{fmla}.   We then generate a
sequence of events consisting of message transmissions from $A$
to $V$ as well as heartbeats at $V$ (when $V$
updates the scores of its peers). The shape of the events we generate
is described by the regex \(\events :=(\msgs\ \hbme)^+\) where 
  \(\msgs := M_1^b \ldots M_i^b\ M_{i+1}^f\ \ldots M_{n}^f\);
  $M_k^l$ denotes sending and receiving $l$ payload messages
from $A$ to $V$ for topic $k$;
$b \in \{0, 1\}$;
$i$ is the number of attacked topics;
$f$ is the number of messages sent for the topics which are not
attacked;
$n = T+2$ is the total number of
topics; and $\hbme$ is a heartbeat event at
$V$.

The event order is unimportant because any permutation of
$\msgs$ between $V$'s heartbeats will have the same effect
on the network. We set $f\textit{=}10$ for each topic
as according to the \Eth\ \gs\ parameters, under normal operation,
this is at least 10\% of the expected mesh message deliveries per
topic, and sending more than $f$ messages can never decrease the score
assigned to an attacker by the victim.

For the \emph{throttling
attack}, the attack reduces the mesh message transmission rate in the
attacked topics to below the threshold set by the \Eth\ parameters
by setting $b\textit{=}1$. For the \emph{blocking attacks},
mesh message transmission is blocked for all of the
attacked topics by setting $b\textit{=}0$. We validate the attacks by
checking that Prop.~1 is eventually always violated by the output
traces of our experiments.

To test our model's performance we ran our
experiments on 100,000 events. Processing one event
generates a cascade of others because when $V$
receives a message, it forwards it to its neighbors, who then forward it
to theirs and so on. Tab.~\ref{tab:eth-attack-times} shows the
time needed to simulate our attacks.

\begin{table}
\begin{center}
\begin{tabular}{|l|c|c|c|c|}
  \hline
  \multirow{2}{*}{Network} &
  \multirow{1}{*}{Throttling} &
  \multicolumn{3}{c|}{Blocking}\\
  \cline{2-5} &
  \multirow{1}{*}{\ag{1}} & \ag{1} & \ag{2} & \ag{3} \\
  \hline
  Ropsten  & 1.3 & 1.3 & 1.3 & 1.3  \\ \hline
  Goerli   & 1.3 & 1.4 & 1.8 & 1.1    \\ \hline
  Rinkeby  & 1.6 & 1.9 & 2.0 & 2.1    \\ \hline
\end{tabular}
\end{center}
\caption{Minutes taken to simulate each attack scenario on each network topology, on a 16GB M1 Macbook Air.}
\label{tab:eth-attack-times}
\end{table}

We observed that for all experiments, the first violation of Prop.~1 occurs 
right after the activation period
(an \Eth\ parameter) has passed. Hence, an attacker peer can start its
attack quickly after it joins a mesh. Experimentally, we observed that
this attack is not transient as attack scores (assigned by victims)
eventually converge to a positive number that stays the same in
successive heartbeats at $V$. By induction, this establishes that our
attacks are perpetual. 
Simulation times for the remaining attacks are similar
and for brevity, we only simulate these attacks until stability
is achieved, which never takes more than $5$ seconds.
Note that the results of these experiments apply equally to real networks, but we did not spin up a real network.  The time taken to execute each attack on a real network will likely differ from the simulation times listed in Tab.~\ref{tab:eth-attack-times}.

\noindent$\bullet$~\textbf{Eclipse Attacks.} We use attack gadgets to construct 
eclipse attacks by just instantiating an attack gadget per neighbor of
a victim such that if they collude, they can target and completely
isolate the victim \ie, the victim will never receive any messages in
the $i$ attacked topics. We tested this attack in the Ropsten topology
by identifying a victim node with four neighboring peers (about 12\%
nodes have degree less than 5),
 and
instantiating its neighbors as attackers using \ag{3} gadgets. We
verified that the victim's message cache contained no message received
in any of the attacked topics while containing messages received in
non-attacked topics, that the attackers were continuously assigned
positive scores by the victim, and that this behavior was perpetual.

\noindent$\bullet$~\textbf{Partition Attacks.}  Given a network graph $G=\langle
V, E\rangle$ and set of victims $S$, we want to identify a set $X$,
preferably of minimal cardinality, such that $X$ is a \emph{vertex
cut} of $G$ that partitions $G$ into disconnected components $\{ S, V
\setminus\{S \cup X\}\}$.  The elements of $X$ are the misbehaving
peers, \ie, each peer in $X$ attacks all of its non-$X$ neighbors,
using our attack gadgets to block the
attacked topics. Hence, no message in an attacked topic can be sent to
a peer in $S$ from a peer outside of the partition and vice versa.  
Finding minimal vertex cuts is NP-hard~\cite{BONSMA2010261}, and can be reduced to either a
Pseudo-Boolean or 0-1 Integer Linear Programming problem.
We synthesized and evaluated partition attacks using the Ropsten
topology by selecting a set of victims~$S$, where $|S|=6$, finding a
minimal vertex cut~$X$, where  $|X| = 2$, and creating the
appropriate attack gadgets.  We verified that each of the victims
did not receive messages in any of the attacked topics that
originated outside of $S$, but did in non-attacked
topics received from outside of $S$; that victim nodes continuously
assigned positive scores to their attackers;  and
that this behavior was maintained forever.


\section{Related Work}
\label{sec:relwork}

\textbf{Attack Discovery.}
\textsc{ProVerif} is an automatic cryptographic protocol verifier based on \textsc{Prolog}
    that can automatically generate attacks against confidentiality and privacy~\cite{blanchet2001efficient}.
\textsc{Tamarin} is similar to \textsc{ProVerif},
and was used to find attacks against \textsc{NAXOS}~\cite{schmidt2012automated},
    \textsc{5G AKA}~\cite{5g_fm_ccs2018},
    the IEEE 802.11 4-way handshake~\cite{singh2020modelling}.
Von Hippel~et.~al. reduced the attacker synthesis problem for protocols
to an LTL model checking problem, and implemented their approach in an 
open-source tool called~\textsc{Korg}~\cite{korg},
which they applied 
to TCP and DCCP~\cite{rfcnlp}.

{\bf Distributed Systems.}
Multiple works modeled and proved theorems about
\textsc{Chord}~\cite{bakhshi2007verification,brunel2018analyzing}.
Woo~et.~al. verified 90 properties of the \textsc{RAFT}~protocol
    using \textsc{Verdi}, a tool they built in the \textsc{Coq} proof assistant~\cite{woos2016planning}.
Although they did not build an executable model, their framework can generate an implementation~\cite{wilcox2015verdi}.
Certain distributed systems might require 
    formally-verified code at every level of the stack.
    Such systems could, \eg, be implemented on top of
    \textsc{sel4}: a high-performance microkernel
    that was verified against an abstract specification using higher-order logic~\cite{klein2009sel4}.

Lamport's modeling language \textsc{TLA+}~\cite{lamport2002specifying}
and the corresponding TLC model checker~\cite{yu1999model} have been
used to analyze properties of distributed systems including
\textsc{Disk Paxos}~\cite{gafni2003disk},
\textsc{MongoRaftReconfig}~\cite{schultz2022formal}, Byzantine
\textsc{Paxos}~\cite{lamport2011byzantizing},
\textsc{Spire}~\cite{koutanov2021spire}, etc.
\textsc{TLA}-style state-machine refinement and Hoare-logic
verification are combined in \textsc{IronFleet}, which was used
to verify a \textsc{Paxos}-based library and a sharded key-value
store~\cite{hawblitzel2015ironfleet}.  The \textsc{Unity}~\cite{unity}
computational model, specification language, and proof system was
successfully applied to numerous distributed systems including a
synchronization scheme for multi-process
handshakes~\cite{park1990distributed}, Segall’s PIF
algorithm~\cite{hesselink1997mechanical}, distributed sorting
algorithms~\cite{bonakdarpour2012framework}, and the Omega Failure
Detector~\cite{bramas2019packet}.  
FM was applied to blockchain protocols in multiple works~\cite{grundmann2022verifying,clockwork,tolmach2021survey}.
In industry, 
Amazon uses a light-weight FM approach to
validate new features 
in their key-value storage
node, \textsc{ShardStore}~\cite{bornholt2021using}.

\textbf{Network Protocols.}
McMillan and Zuck applied specification-based testing to the \textsc{Quic} protocol,
    and found multiple implementation errors, some of which caused vulnerabilities~\cite{quic_fm_sigcomm19}.
Wu~et.~al. formally modeled the Bluetooth stack using \textsc{ProVerif}, and found five known vulnerabilities and two new ones~\cite{bluetooth_fm_sp2022}.
Chothia~et.~al. showed how \textsc{ProVerif} can be used
to verify distance-bounding protocols, \eg\ those 
used by MasterCard and NXP~\cite{chothia2018modelling}.
Chothia modeled the \textsc{MUTE} anonymous file-sharing system
using the $\pi$-calculus, and proved the system insecure~\cite{chothia2006analysing}.
Cremers~et.~al. modeled all handshake modes of TLS~1.3 using \textsc{Tamarin}, and discovered an unexpected behavior~\cite{tls13_fm_ccs2017}.
Although most of these use model checking \emph{or} theorem proving,
    Manolios~et.~al. link the two 
    to verify
    ABP~\cite{bisimulation1999linking}.

\section{Discussion}
Our work highlights three concrete steps that developers can take to harden \gs\ and other similar systems.  First, they can formalize the properties the protocol is designed to satisfy (the protocol goals) and the protocol requirements (\eg, that weights should be non-zero).  Simply formalizing properties and requirements enables light-weight FM, and assures developers know when they can rely on the protocol and for what.  Second, they can design a score that does not use caps, to avoid the vulnerability we reported in which above a certain score, misbehavior goes unreported.  Third, they can leverage model-based counterexample generation to test new protocol configurations before deploying to apps like \Eth\ or \fc.  This can be done for \gs\ using our code, or for other protocols by adapting the techniques laid out in this work.
For \fc\ or \Eth, one simply needs to update the twp values to model the new configuration, and our system will assess it.

Our work illustrates how heuristic ``defenses'' enable attacks that exploit their edge-cases, implying protocol designers should design defense mechanisms from first principles, or leverage FM to rule out such edge cases.  Unfortunately, there are too few FM tools for analyzing the security of protocols and distributed systems at scale, and existing ones are too difficult to use.  Multiple reviews found that security practitioners prioritize ease-of-use when choosing an FM tool, \eg, choosing a model-checker (which cannot scale but are easy to use) over a theorem prover (which can scale but is difficult to use)~\cite{clarke1996formal,kulik2022survey}.  An important research direction is thus the translation of cutting-edge FM tools (e.g.~\cite{jaber2021quicksilver}) to software that can be easily used by non-expert developers of protocols and distributed systems.

\section{Conclusion}
\label{sec:concl}
In this paper we rigorously studied \gs~and its security
using the \acls~theorem prover.  We created a complete
model of the protocol and formalized security properties based on
the prose \gs~specification. We showed that the properties depend on
how the \score\ is
  configured. Of two
well-known applications, \fc~and \Eth, only \fc~satisfied all of our
properties.  We showed that on any \Eth\
network, of any topology and size,
we can synthesize attacks
where certain peers continuously misbehave by never
forwarding topic messages, but are never
identified as misbehaving and thus
are never pruned from the network.
We ethically disclosed our results to Protocol Labs
  and the Ethereum Foundation, who agreed with our findings.  In
  addition, the \gs\ developers at Protocol Labs publicly endorsed our
  model as a formal specification for \gs.

Writing this model
required effort comparable to implementing the protocol in a
programming language, while
providing a formal model that we can reason about with mathematical
precision.  We did not have to use extensive manual testing because
using \acls\ to analyze properties helped us find attacks.  Our
work required less manual effort than that expended by
the \gs\ developers, and found attacks that they missed.
  Developers
  interested in applying our approach to their gossip protocols can
  build on our formalization, rather than starting from scratch.

{\small \noindent~\textbf{Acknowledgements}~We sincerely thank
Yiannis Psaras, Dimitris Vyzovitis, Nishant Das, Adrian Manning, Max
Inden, and others at Protocol Labs and the Ethereum Foundation for
their valuable feedback.}

\bibliographystyle{IEEEtran}
\bibliography{main.bib,p2p.bib}

\appendices
\section{Implementation Details in \acls}
Consider the following excerpts from our model.
\begin{lstlisting}[language=Lisp]
(defdata pos-rat (range rational (0 <= _)))
(defdata topic-counters
  (record (invalidMessageDeliveries . pos-rat)
    (meshMessageDeliveries    . pos-rat)
    (meshTime                 . pos-rat)
    (firstMessageDeliveries   . pos-rat)
    (meshFailurePenalty       . pos-rat)))
\end{lstlisting}
We define \tctrs, a named record to store topic based
performance counters. Since these counters can never be negative, and
may be rational (due to decay), we specify their types as appropriately
defined \texttt{pos-rat}s. We then define a map \ptt\ to store \tctrs\
per peer per topic as follows:
\begin{lstlisting}[language=Lisp]
(defdata pt (cons peer topic))
(defdata pt-topic-counters-map (alistof pt topic-counters))
\end{lstlisting}
Now we need a lookup function to find \tctrs, given a peer and a
topic.
\begin{lstlisting}[language=Lisp]
(definec lookup-topic-counters (p :peer top :topic map :pt-topic-counters-map) :topic-counters
  (match map
    (() (new-topic-counters))
    ((((!p . !top) . tct) . &) tct)
    ((& . rst) (lookup-topic-counters p top rst))))
\end{lstlisting}

We use \texttt{match} to pattern-match \texttt{map} against possible
syntactic structures. If \texttt{map} is empty, we return a
new \tctrs, with all counters initialized to zero. Else, if the
pair of peer and topic exactly matches the key in the first
key-value pair of \texttt{map}, we return the corresponding value,
otherwise we recurse on the rest of \texttt{map}.

Upon admitting \texttt{lookup-topic-counters}, \acls\ extends its logic with
(1) a definitional axiom: given input arguments satisfy their types,
calling \texttt{lookup-topic-counters} equals its function body, and (2) a
function contract theorem: given input arguments satisfy their types
calling \texttt{lookup-topic-counters} returns a \tctrs, as specified by the
function output type. Such axioms could introduce unsoundness if
\texttt{lookup-topic-counters} did not terminate. So before admitting the
function, \acls\ uses termination analysis to prove that
\texttt{lookup-topic-counters} is indeed terminating. Hence, admitting
\texttt{lookup-topic-counters} produces theorems about its definition,
termination and I/O contracts.

\begin{table}
\begin{center}
\begin{tabular}{ |l|l|l| } \hline Constant or Weight & \texttt{MESSAGES}
  & \texttt{BLOCKS} \\ \hline
  \texttt{TopicWeight} & 1 & 1 \\
  \texttt{TopicCap} & $0$ & $0$ \\
  $w_1$(t) & 2.78 & 0.027 \\
 $w_2(t)$ & 0.5 & 5 \\
 $w_3(t)$ & 0 & 0 \\
 $w_{3b}(t)$ & 0 & 0 \\
 $w_4(t)$ & -1000 & -1000 \\
 $w_5$ (global) & 1 & 1 \\
 $w_6$ (global) & -100 & -100\\
 $w_7$ (global) & -10 & -10 \\ 
 \texttt{D} & 8 & 8 \\\hline
\end{tabular}
\end{center}
\caption{\fc's \texttt{twp}.  Adapted from \url{github.com/filecoin-project/lotus}.}
\label{tab:fctwp}
\end{table}

\begin{table}
  \begin{center}
    \begin{tabular}{|l|l|l|l|l|l|}
      \hline
      Constant or Weight & \BLOCKS & \AGG & \SUBONE & \SUBTWO & \SUBTHREE \\ \hline
      \texttt{TopicWeight} & 0.8 & 0.5 & 0.33 & 0.33& 0.33\\ 
      \texttt{TopicCap} & 32.72 & 32.72 & 32.72 & 32.72 & 32.72 \\ 
      $w_1(t)$ & 0.0324 & 0.0324& 0.0324 & 0.0324 & 0.0324 \\ 
      $w_2(t)$ & 1 & 0.128 & 0.95 & 0.95 & 0.95 \\ 
      $w_3(t)$ & -0.717 & -0.064& -37.55 & -37.55 & -37.55 \\ 
      $w_{3b}(t)$ & -0.717 & -0.064& -37.55 & -37.55 & -37.55 \\ 
      $w_4(t)$  & -140.45 & -140.45 & -4544 & -4544 & -4544 \\ 
      $w_5$ (global) & 1 & 1 & 1 & 1 & 1 \\ 
      $w_6$ (global) & -35.11   & -35.11& -35.11 & -35.11 & -35.11 \\ 
      $w_7$ (global) & -15.92   & -15.92& -15.92 & -15.92 & -15.92 \\ 
      \texttt{D} & 8 & 8 & 8 & 8 & 8 \\
      \hline
    \end{tabular}
  \end{center}
  \caption{\Eth's \texttt{twp}.  Adapted from \url{github.com/silesiacoin/prysm-spike}.}
  \label{tab:ethtwp}
\end{table}

\begin{table}
\begin{center}
\begin{tabular}{ |l|l|l|l| } \hline Constant or Weight & Pathological 1-5 & Good  1 & Good 2 \\\hline
\texttt{TopicWeight} & 40 & 0.5 & 0.5 \\
\texttt{TopicCap} & 5 & 100 & 10 \\ 
  $w_1(t)$ & 10 & 0.027 & 0.027 \\ 
  $w_2(t)$ & 10 & 5 & 5 \\ 
  $w_3(t)$ & -1  & -1000 & -1000 \\ 
  $w_{3b}(t)$ & -1 & -1000 & -1000 \\ 
  $w_4(t)$ & -1  & -1000 & -1000 \\
  $w_5$ (global) & 10  & 1 & 1 \\ 
  $w_6$ (global) & -1 & -100   & -100 \\
  $w_7$ (global) & -1 & 10   & 10 \\  
  \texttt{D} & 5 & 8 & 8 \\\hline
\end{tabular}
\end{center}
\caption{Our pathological \twp\ consists of five topics all configured per column 2.  Our good configuration \twp\ consists of two topics, given in columns 3 and 4, and satisfies all our properties.}
\label{tab:pathtwp}
\end{table}

\begin{figure}
    \centering
    \centering
\begin{tikzpicture}
\draw[draw=black] (0, 0) rectangle ++(2, 3.8);
\node[] (gr) at (1,3.4) {\texttt{gp}};
\node[draw=black] (ps1) at (1,2.8) {\texttt{ps1}};
\node[] (ps2) at (1,2.2) {\dots};
\node[draw=black] (psK) at (1,1.6) {\texttt{psK}};
\node[] (dots) at (1,1) {\dots};
\node[draw=black] (psN) at (1,0.4) {\texttt{psN}};
\node[draw=black,align=left] (init) at (1,-1) {\texttt{init-}\\\texttt{evnts}};
\node[draw=black,align=left] (eventsweaver) at (4,-1) {\texttt{events-}\\\texttt{weaver}};
\draw[-latex] (init) to (eventsweaver);
\node[draw=black] (peer-trx) at (4,1.6) {\texttt{ps-trx}};
\draw[-latex] (psK) to (peer-trx);
\draw[-latex] (eventsweaver) to (peer-trx);
\draw[draw=black] (6, 0) rectangle ++(2, 3.8);
\node[] (grp) at (7,3.4) {\texttt{gp}'};
\node[draw] (ps1c) at (7,2.8) {\texttt{ps1}};
\node[] (ps2p) at (7,2.2) {\dots};
\node[draw=black] (psKp) at (7,1.6) {\texttt{psK}'};
\node[] (dots) at (7,1) {\dots};
\node[draw=black] (psNc) at (7,0.4) {\texttt{psN}};
\draw[-latex] (peer-trx) to (psKp);
%
%
\node[draw=black] (evnt) at (7,-1) {\texttt{evnt}};
\draw[-] (4,0.9) -- (5.5,0.9) -- (5.5,-1);
\draw[-latex] (5.5,-1) to (evnt);
\node[draw=black,fill=white] (car) at (4,0.5) {\texttt{car}};
%
\draw[-] (4.55,1.35) to (4.55,1.1);
\draw[-latex] (4.55,0.8) to (4.55,-0.55);
\end{tikzpicture}
    \caption{Data-flow diagram of the \group~transition function,
      \texttt{gs-trx} which takes as input a \group~called
      \texttt{gp}, a list of \evnt s called \texttt{init-evnts}, a
      \twp, and an oracle.  The first \evnt\ coming out of \evnt s
      weaver determines which \ps~\texttt{psK} in the
      \group~\texttt{gp} will get updated this round.  The \twp~and
      oracle are both passed into the \ps~transition function calls,
      and are
      omitted for simplicity. The \texttt{events-weaver} function
      splices the \texttt{init-evnts} with the events emitted by the
      most recent previous application of the \texttt{ps-trx}
      function.  The \texttt{car} function selects the first
      \evnt~from the list of \evnt~s generated by
      \texttt{events-weaver}, which serves as input to \texttt{ps-trx}
      in this round, as well as an output of the overall
      \group~transition function.}
    \label{fig:group-fsm}
\end{figure}

\begin{figure*}
    \centering
    \centering
\begin{tikzpicture}
\draw[draw=black] (0, 0) rectangle ++(2, 3.8);
\node[] (ps) at (1,3.4) {\texttt{ps}};
\node[draw=black] (nts) at (1,2.8) {\texttt{nts}};
\node[draw=black] (ms) at (1,2.2) {\texttt{ms}};
\node[draw=black] (tctrs) at (1,1.6) {\texttt{tctrs}};
\node[draw=black] (gctrs) at (1,1) {\texttt{gctrs}};
\node[draw=black] (scores) at (1,0.4) {\texttt{scores}};
\node[draw=black,minimum size=2cm] (nts-trx) at (4.2, 2.3) {\texttt{nts-trx}};
\draw[-latex] (nts) to (3.2,2.8);
\draw[-] (tctrs) -- (2.5,1.6) -- (2.5,2.2);
\draw[-latex] (2.5,2.2) to (3.2,2.2);
\draw[-] (gctrs) -- (2.7,1) -- (2.7,1.9);
\draw[-latex] (2.7,1.9) to (3.2,1.9);
\draw[-] (scores) -- (2.9,0.4) -- (2.9,1.6);
\draw[-latex] (2.9,1.6) to (3.2,1.6);
\node[draw=black,minimum size=1cm] (append) at (10.2,-1.3) {\texttt{append}};
\node[draw=black,align=left] (init) at (1,-1.5) {\texttt{evnt}};
\draw[-] (init) -- (2.3,-1.5)
         -- (2.3,2.5);
\draw[-latex] (2.3,2.5) to (3.2,2.5);
\draw[draw=black] (6.2, 0) rectangle ++(2, 3.8);
\node[] (ps-p) at (7.2,3.4) {\texttt{ps}'};
\node[draw=black] (nts-p) at (7.2,2.8) {\texttt{nts}'};
\node[draw=black] (ms-copied) at (7.2,2.2) {\texttt{ms}};
\node[draw=black] (tctrs-p) at (7.2,1.6) {\texttt{tctrs}'};
\node[draw=black] (gctrs-p) at (7.2,1) {\texttt{gctrs}'};
\node[draw=black] (scores-p) at (7.2,0.4) {\texttt{scores}'};
\draw[-stealth] (5.2,2.8) to (nts-p);
\draw[-] (5.2,2.2) -- (6,2.2) -- (6,1.6);
\draw[-latex] (6,1.6) to (tctrs-p);
\draw[-] (5.2,1.9) -- (5.8,1.9) -- (5.8,1);
\draw[-latex] (5.8,1) to (gctrs-p);
\draw[-] (5.2,1.6) -- (5.6,1.6) -- (5.6,0.4);
\draw[-latex] (5.6,0.4) to (scores-p);
\node[draw=black,minimum size=1.4cm] (mstrx) at (10.2,1.6) {\texttt{ms-trx}};
\draw[-latex] (ms-copied) to (9.45,2.2);
\draw[-] (gctrs-p) -- (9,1) -- (9,1.2);
\draw[-latex] (9,1.2) to (9.45,1.2);
\draw[-] (2.3,-0.3) -- (3.9,-0.3);
\draw[-] (4.6,-0.3)
    -- (8.7,-0.3)
    -- (8.7,1.9);
\draw[-latex] (8.7,1.9) to (9.45,1.9);
\draw[-latex] (tctrs-p) to (9.45,1.6);
\draw[-] (nts-trx) to
    (4.2,-1.3);
\draw[-latex] (4.2,-1.3) to (append);
%
\draw[-latex] (mstrx) to (append);
\draw[draw=black] (12.2, 0) rectangle ++(2, 3.8);
\node[] (ps-pp) at (13.2,3.4) {\texttt{ps}''};
\node[draw=black] (nts-pc) at (13.2,2.8) {\texttt{nts}'};
\node[draw=black] (ms-p) at (13.2,2.2) {\texttt{ms}'};
\node[draw=black] (tctrs-pp) at (13.2,1.6) {\texttt{tctrs}''};
\node[draw=black] (gctrs-pp) at (13.2,1) {\texttt{gctrs}''};
\node[draw=black] (scores-pc) at (13.2,0.4) {\texttt{scores}'};
\draw[-latex] (10.95,2.2) to (12.8,2.2);
\draw[-] (10.95,1.9) -- (12,1.9)
         -- (12,1.6);
\draw[-latex] (12,1.6) to (12.5,1.6);
\draw[-] (10.95,1.6) -- (11.8,1.6)
         -- (11.8,1);
\draw[-latex] (11.8,1) to (12.5,1);
\node[draw=black] (output) at (13,-1.3) {\texttt{evnts}};
\draw[-latex] (append) to (output);

\end{tikzpicture}
    \caption{Data-flow diagram of the \ps~transition function,
      \texttt{ps-trx}. The function takes as input a \ps~called
      \texttt{ps}, a list of \evnt s called \texttt{evnt}, a \twp, and
      an oracle.  The oracle is used to model nondeterministic
      decisions, \eg, the specific subset of peers to send gossip
      to. Both the \twp~and the oracle are omitted from the
      diagram. 
      The function outputs a new \ps~called \texttt{ps}', and a
       list of \evnt s called \texttt{evnts'}.  The box labeled \texttt{nts-trx}
      denotes a transition function for the \nts, which outputs an
      updated \nts~called \texttt{nts}' as well as a
      list of emitted \evnt s.  It also generates new \tctrs~and
      \gctrs, and calculates peer scores. Next, the \mst~component
      \texttt{ms}' of the \ps~\texttt{ps}' is fed into the
      \texttt{ms-trx} function, along with the updated \tctrs~and
      \gctrs, and the initial \evnt.  The final \ps~\texttt{ps}''
      consists of \texttt{nts}', \texttt{ms}', \texttt{tctrs}'',
      \texttt{gctrs}'', and \texttt{scores}'.  A list
      of events called~\texttt{evnts} is also emitted by splicing the
      lists emitted by the \texttt{nts-trx} and \texttt{ms-trx}
      functions.}
    \label{fig:peer-state-fsm}
\end{figure*}

\begin{table*}
\begin{center}
\begin{tabular}{ |l|l|p{0.35\linewidth}|l| } \hline 
Parameter & Type & Description & Guidance \\ \hline 
\texttt{PruneBackoff} & Duration & Duration before pruned peer may re-graft & Default to 1 minute \\
\texttt{UnsubscribeBackoff} & Duration & Duration before unsubscribed peer may re-subscribe & Default to 10 seconds \\
\texttt{FloodPublish} & Boolean & Enable/disable optional flood publishing & Default to true \\
\texttt{GossipFactor} & Float & Fraction of positive-scoring peers to emit gossip to & Default to 0.25, must be in [0, 1] \\
\texttt{D} & Integer & Desired outbound degree of each mesh & Default to 6 \\
\texttt{Dlow} & Integer & Lower bound for outbound degree of each mesh & Default to 4 \\ 
\texttt{Dhi} & Integer & Upper bound for outbound degree of each mesh & Default to 12 \\ 
\texttt{Dlazy} & Integer & Desired outbound degree for gossip emission & Default to \texttt{D} \\
\texttt{HeartbeatInterval} & Duration & Duration between \hbm s & Default to 1 second \\
\texttt{FanoutTTL} & Duration & Time-to-live for fanouts & Default to 1 minute \\
\texttt{SeenTTL} & Duration & Time-to-live for cache of seen message identifiers & Default to 2 minutes \\
\texttt{McacheLen} & Integer & Number of history windows in message cache & Default to 5 \\
\texttt{McacheGossip} & Integer & Number of history windows to use when emitting gossip & Default to 3 \\
\texttt{Dscore} & Integer & Number of highest-scoring peers to retain when pruning due to over-subscription & 4 or 5 for a \texttt{D} of 6 \\
\texttt{Dout} & Integer & Number of outbound connections to keep in a mesh & Default to 2 for \texttt{D}=6, must be in [\texttt{Dlo}, \texttt{D}/2] \\
\texttt{GossipThreshold} & Float & Only emit gossip to peers who score above this threshold & Must be $< 0$ \\
\texttt{PublishThreshold} & Float & Only send new messages to peers who score above this threshold & Must be $\leq \texttt{GossipThreshold}$ \\
\texttt{GraylistThreshold} & Float & Ignore control messages from peers scoring below this threshold & Must be $<$ \texttt{PublishThreshold} \\
\texttt{OpportunisticGraftThreshold} & Float & Opportunistic grafting is triggered when the median score of neighbors in a mesh falls below this threshold & Must be $\geq 0$ \\
\texttt{DecayInterval} & Duration & Interval at which counters decay & \\
\texttt{DecayToZero} & Float & When \inds~fall below this value they round down to zero & Should be close to 0.0 \\
\texttt{RetainScore} & Duration & Duration to retain peer scores after they disconnect & \\
\hline
\end{tabular}
\end{center}
\caption{Parameters used by \gs's defense mechanisms.  Descriptions adapted from the prose specification~\cite{gs1.0,gs1.1}.}
\label{tab:defaultParams}
\end{table*}

\begin{figure}
\begin{lstlisting}[language=Lisp]
(property (ptc :pt-tctrs-map pcm :p-gctrs-map p :peer top :topic)
	  :hyps  (^ (member-equal (cons p top) (acl2::alist-keys ptc))
                    (> (lookup-score p (calc-nbr-scores-map ptc pcm *eth-twp*)) 0))
          (> (calcScoreTopic (lookup-tctrs p top ptc) (mget top *eth-twp*))
             0))
\end{lstlisting}
\caption{Property 1 definition in \acls\ for \Eth.}
\label{lst::p1}
\end{figure}

\begin{figure}
\begin{lstlisting}[language=Lisp]
(property (ptc :pt-tctrs-map pglb :p-gctrs-map p :peer top :topic delta-p3 :non-neg-rational
       delta-p3b :non-neg-rational delta-p4 :non-neg-rational delta-p6 :non-neg-rational 
       delta-p7 :non-neg-rational)
:hyps (^ (member-equal top (strip-cars *eth-twp*))
   (member-equal (cons p top) (strip-cars ptc))
         (member-equal p (strip-cars pglb))
   (> (+ delta-p3 delta-p3b delta-p4 delta-p6 delta-p7) 0))
(b* ((tc (lookup-tctrs p top ptc))
     (glb (lookup-gctrs p pglb))
     (new-tc (update-meshMessageDeliveries
    tc
    (- (tctrs-meshMessageDeliveries tc) delta-p3)))
     (new-ptc (put-assoc-equal `(,p . ,top) new-tc ptc)))
  (> (lookup-score p (calc-nbr-scores-map ptc pglb *eth-twp*))
     (lookup-score p (calc-nbr-scores-map new-ptc pglb *eth-twp*)))))
\end{lstlisting}
\caption{Property 2 definition in \acls, for \Eth.}
\label{lst::p2}
\end{figure}

\begin{figure}
\begin{lstlisting}[language=Lisp]
(property
 (imd mmd mt fmd mfp p :non-neg-rational wtpm :wp)
 (=> (^ (== wtpm (cdr (assoc-equal 'BLOCKS *ETH-TWP*)))
        (>= (params-meshMessageDeliveriesCap (cdr wtpm))
            (params-meshMessageDeliveriesThreshold (cdr wtpm)))
        (> mt (params-activationWindow (cdr wtpm))))
     (>= (calcScoreTopic (tctrs imd mmd  (+ p mt) fmd mfp) wtpm)
         (calcScoreTopic (tctrs imd mmd  mt fmd mfp) wtpm))))
\end{lstlisting}
\caption{Property 3 definition in \acls\ for \Eth, for the \texttt{BLOCKS} topic.  Analogous properties are written for each other topic.  Property 4 is checked automatically by \acls\ thus does not need to be explicitly written down.}
\label{lst::p3}
\end{figure}

\section{Use of \gs\ in Applications}
\gs~is used most notably in \fc\ and \Eth. 
\fc~is a decentralized storage solution based on Proof-of-Space-Time.
It is 
a P2P alternative to the client-server model,
    where content (\eg, websites) are addressed by their hashes
    and nodes can earn cryptocurrency by acting as hosts.
Peers wishing to publish content 
    find hosts using \emph{ask} orders in a distributed auction house
    (the hosts respond with \emph{bid} orders).
Both types of orders are disseminated using \gs. 
\fc~uses two topics: \texttt{BLOCKS} and \texttt{MSGS}~\cite{FCwhitepaper}.
In 2021, \fc\ had $>$14 EiB in network storage capacity, $>$3,600 network storage providers, and tens of millions of uploads by tens of thousands of users~\cite{fcStats}.  Many real-world applications are built on top of \fc\ including the Inter-Planetary File System (IPFS), various Non-Fungible Token (NFT) marketplaces, etc.

\Eth\ is the second most valuable cryptocurrency, after 
Bitcoin.
    \Eth\ supports Turing-complete smart contracts with 
    fine-grained control over the amount of value being exchanged,
    an internal program state, and
    access to blockchain data such as nonces~\cite{EthWhitepaper}.
Over 2900 applications are built on \Eth, some with millions of active daily users, including NFTs, Decentralized Autonomous Organizations (DAOs), Decentralized Finance apps (DeFi), etc.~\cite{ethApps,ethAppArticle}
\gs\ is the primary messaging layer protocol in \Eth.  
It is used to disseminate all kinds of data throughout the chain,
    including newly signed blocks,
    attestations,
    payload encodings, over dozens of topics~\cite{EthGsUsage}.

\section{Properties.} They are defined in Alg. \ref{lst::p1}, \ref{lst::p2}, \ref{lst::p3}.

\end{document}